\documentclass[twocolumn,trackchanges]{aastex701}

\begin{document}

\title{Aligned, Misaligned and Polar Orbits of Hot Jupiters: Measuring Spin-Orbit Angles via Doppler Tomography with HARPS-N} 

\author[gname='Zuzana',sname='Balkóová']{Z. Balkóová}
\affiliation{Astronomical Institute of the Czech Academy of Sciences, Fričova 298, 251~65 Ondřejov, Czech Republic}
\affiliation{Astronomical Institute of Charles University, V Holešovičkách 2, 180~00 Prague, Czech Republic}
\email{fakeemail1@google.com}  

\author[gname=Jiří, sname='Žák']{J. Žák} 
\affiliation{Astronomical Institute of the Czech Academy of Sciences, Fričova 298, 251~65 Ondřejov, Czech Republic}
\email{fakeemail2@google.com}

\author[gname=Marek,sname=Skarka]{M. Skarka}
\affiliation{Astronomical Institute of the Czech Academy of Sciences, Fričova 298, 251~65 Ondřejov, Czech Republic}
\email{fakeemail3@google.com}

\author[gname=Emil,sname=Knudstrup]{E. Knudstrup}
\affiliation{Department of Space, Earth and Environment, Chalmers University of Technology, 412~93 Gothenburg, Sweden}
\email{fakeemail4@google.com}

\author[gname=Pavol,sname=Gajdoš]{P. Gajdoš}
\affiliation{Astronomical Institute of the Czech Academy of Sciences, Fričova 298, 251~65 Ondřejov, Czech Republic}
\affiliation{Institute of Physics, Faculty of Science, Pavol Jozef \v{S}af\'arik University, Park Angelinum 9, 040 01 Ko\v{s}ice, Slovakia}
\email{fakeemail5@google.com}

\author[gname=Andrea,sname=Bignamini]{A. Bignamini}
\affiliation{INAF – Osservatorio Astronomico di Trieste, via Tiepolo 11, 34143 Trieste, Italy}
\email{fakeemail6@google.com}

\author[gname=Petr,sname=Kabáth]{P. Kabáth}
\affiliation{Astronomical Institute of the Czech Academy of Sciences, Fričova 298, 251~65 Ondřejov, Czech Republic}
\email{fakeemail7@google.com}

\begin{abstract} 
Although the migration of hot Jupiters is not yet fully understood, measurements of the projected spin–orbit angle $\lambda$ help shed light on the processes involved. Here we present Doppler tomography of three known hot Jupiters to determine their $\lambda$ orientation: HAT-P-49\,b, HAT-P-57\,A\,b, and XO-3\,A\,b. Our analysis explores the impact of cross-correlation processing methods on the detectability and characterization of the planet's Doppler shadow using up to three independent routines for cross-correlation functions extraction; those being: \texttt{Yabi}, \texttt{iSpec} and \texttt{IRAF}. After accounting for differences among the results obtained with the various routines, we report: first, the HAT-P-49 system is a case of a hot Jupiter on a polar orbit with $\lambda=-85.3\pm1.7\,^\circ$, second, HAT-P-57\,A indicates practically no deviation of the planet's projected orbit from the host's equatorial plane with $\lambda=-0.4_{-1.9}^{+1.4}\,^\circ$, and third, the XO-3\,A system with the measured value of $\lambda=38_{-4}^{+3}\,^\circ$ lies in between an aligned and a perpendicular orientation, which is a less populated region of the spin-orbit distribution. Our findings highlight both the diversity of spin–orbit angles among close-in giant planets and the potential discrepancies in their measurement that can arise from different approaches to constructing the cross-correlation functions.

\end{abstract}



\section{Introduction}
While the relative contributions of various formation and migration scenarios of hot Jupiters remain uncertain, they are commonly grouped into three main categories: in situ formation under extreme conditions \citep[e.g.][]{Batygin2016, Dawson2018}, disk migration with torques prominently shrinking the planet's semimajor axis \citep[e.g.][]{Lin1986, Murray1998}, or high-eccentricity migration (HEM) driven by perturbation of the planet by one or more massive bodies \citep[e.g.][]{Fabrycky2007, Giacalone2017}), followed by tidal dissipation of its orbital energy and further circularization of its orbit while undergoing substantial oscillations in inclination. An alternative to the HEM scenario is the idea that stellar obliquities are primordial and any misalignment arises due to chaotic accretion, cluster dynamics, or magnetic interactions \citep[e.g.][]{Bate2010, Lai2011}. A number of discovered hot Jupiters exhibit high values of \(\lambda\), the projected angle between the planet's orbital plane and the star's spin axis, which suggests that HEM may play a role in some systems. Given that realignment of such perturbed orbits can take several billion years \citep{Albrecht2012}, it is plausible that we observe some of these systems during this prolonged transitional phase. Furthermore, \cite{Winn2010} first discovered that stars above the Kraft break \citep[$\approx 6250$\,K,][]{Kraft1967} hosting hot Jupiters generally tend to have high obliquities, suggesting that tidal realignment is more effective in stars with convective envelopes. Later, \cite{Albrecht2012} and \cite{Rice2022} showed that although hot Jupiters on circular orbits around hot stars exhibit a wider range of spin-orbit angles, no such temperature-dependent trend is seen among the eccentric sample of these planets. This implies that their orbits have not yet been circularized or aligned.
\par
The angle \(\lambda\), known as the sky-projected obliquity (or spin-orbit angle), can be found via the Rossiter-McLaughlin (RM) effect \citep{Rossiter1924, McLaughlin1924}, a spectroscopic anomaly detectable during a transit, within which a planet selectively covers different parts of the stellar surface; that is, regions producing different radial velocities towards the observer. This distortion is most commonly measured via the radial velocity (RV)-RM effect, where the deficit of flux in absorption stellar line profiles affects the net stellar radial velocity. 
Although this method has proven reliable, it is not universal, as extracting radial velocities is challenging for hot stars, which have too few spectral lines for precise RV measurements and therefore produce very large errorbars. The situation is even more problematic for rapid rotators, where line broadening further increases the RV uncertainties. In such cases, a technique called Doppler tomography offers a more direct approach \citep{Johnson2014} by tracking the actual distortion within the evolving shape of the line profiles. According to the $\alpha$ ratio 

\begin{equation}
\alpha = \frac{(v \sin i_*) (R_p/R_\star)}{\sqrt{\sigma_{\rm inst}^2 + \sigma_{\rm micro}^2 + \sigma_{\rm macro}^2}},
\end{equation}

presented by \cite{Albrecht2022}, the detectability of the planet's presence in the stellar spectra can be significantly affected not only by the star's projected rotational velocity $v \sin i_*$ and the radii ratio $R_p/R_\star$, but also by the magnitude of micro- and macroturbulences $\sigma_{\rm micro}$ and $\sigma_{\rm macro}$, and the spectrograph's instrumental broadening $\sigma_{\rm inst}$. Moreover, \cite{Brown2012} demonstrated another advantage of Doppler tomography over the RM effect, noting the $v \sin i_*-\lambda$ degeneracy that exists in systems with low impact parameter - that is, when the planet appears to transit very close to the center of the stellar disk. This degeneracy is much weaker when using the tomographic method.
\par
For the first time among exoplanets, \cite{CollierCameron2010A} applied Doppler tomography on HD~189733\,b, and in the same year's paper \citep{CollierCameron2010B} on WASP-33\,b, both being giant gas planets with short orbital periods. After these first applications, more Doppler tomography analyzes followed \citep[e.g.][]{Zhou2016, Albrecht2013}) as an alternative or complement to previous, traditional spin-orbit angle measurements. These, however, can show disagreement, especially when using different data. Such an example is the system WASP-71, measured by \cite{Smith2013} using CORALIE spectrograph via the RV-RM effect to have $\lambda=20.1\pm9.7\,^{\circ}$, compared to the value of $\lambda=-1.9^{+7.1}_{-7.5}\,^{\circ}$ given by \cite{Brown2017} using Doppler tomography on HARPS data. Another example of an RV-RM effect versus Doppler tomography spin-orbit angle measurement discrepancy can be seen in the works of \cite{Sedaghati2022} and \cite{Knudstrup2022}, yielding $\lambda=-42\pm1\,^{\circ}$ and $\lambda=-2\pm6\,^{\circ}$, respectively, for HD~332231. In this case, \cite{Knudstrup2022} also note that using different ephemerides could have amplified the differences between these results.
\par
Thanks to the availability of high-resolution spectroscopic data, the number of spin-orbit measurements have seen a surge in recent years \citep{Bourrier2023, knud24, zak24}. Yet, the underlying mechanisms shaping the spin-orbit distribution remain elusive and more measurements covering the whole parameter space are needed to confirm or rebut tentative trends such as the "Prepondarence of Perpendicular Planets" suggesting certain orbital configurations are preferred rather than the distribution being isotropic \citep{alb21,attia23}.
\par
In this work, we revisit archival spectroscopic data for a sample of stars hosting hot Jupiters — namely HAT-P-49, HAT-P-57\,A, and XO-3\,A — and apply Doppler tomography to investigate their projected spin-orbit angles \(\lambda\). We focus not only on detecting the Doppler shadow in each system, but also on comparing how different approaches to cross-correlation affect the resulting tomographic signals. In particular, we examine the impact of using various synthetic templates and/or line masks in the cross-correlation function (CCF) construction, aiming to assess how these choices influence the measured spin-orbit angles, placing our results in the context of previous spin-orbit angle measurements derived either from Doppler tomography or RV-RM effect. Additionally, we investigate several dynamical timescales that could have influenced and possibly driven the evolution toward the planetary orbits we observe today.

\hfill \break
\section{Data samples}
\begin{table*}[]
    \centering
    \caption{A sample of objects with data available from the TNG archive. Columns denote elemental statistics for each observation. In HAT-P-57\,A, the dates are not sorted chronologically, as this arrangement is more practical for later visualizations.}
    \begin{tabular}{c|c c c c c c}
    Target & Date & Total no. spectra & No. transit spectra & Exp. time & Avg. SNR & Airmass \\
    & [y-m-d] & & & [s] & at 550~nm per pixel & (start-lowest-end) \\
    \hline \hline
    \noalign{\smallskip}
        HAT-P-49 & 2020-07-30 & 126 & 73 & 180 & 22.55 & 1.31-1.00-1.92 \\
        HAT-P-57\,A & 2021-06-02 & 19 & 11 & 1200 & 54.14 & 1.64-1.05-1.32 \\
        & 2019-06-23 & 44 & 20 & 600 & 47.61 & 1.66-1.05-1.88 \\
        & 2019-06-28 & 39 & 21 & 600 & 51.66 & 1.76-1.05-1.40 \\
        XO-3\,A & 2015-11-01 & 54 & 21 & 500 & 45.45 & 1.29-1.14-1.50 \\
    \end{tabular}
    \label{tab:objects_selection}
\end{table*}

In many cases, high-resolution spectroscopic transit observations are obtained with the explicit goal of detecting the RV-RM anomaly, yet the full information content of the line profiles is not always utilized, and Doppler tomography - despite its advantage particularly for rapidly rotating stars - remains a less commonly applied technique. For that, we selected a sample of hot Jupiter systems and searched for publicly available spectroscopic data in the archive of the Telescopio Nazionale Galileo \citep[TNG,][]{TNGarchive}. We found three targets, of which - to our knowledge - only HAT-P-49 had previously been analysed via Doppler tomography using these data. We focused primarily on files containing extracted cross-correlation functions, but also their respective 1D spectra. These could be used for external rework of the CCFs, allowing for direct comparison between different processing pipelines and template choices. Each of our chosen targets has at least one transit observation (see Tab \ref{tab:objects_selection}) with sufficient phase coverage to track the Doppler shadow.
\par
The three systems selected for this study - all coming from archival HARPS-N observations \citep{HARPSN_Cosentino, HARPSN} - were chosen to reflect different scientific and methodological motivations.
\begin{enumerate}
\item HAT-P-49 has already been analyzed using the same spectroscopic dataset; here, we aim to test whether we can reproduce the published results and assess the consistency of spin-orbit angle measurements when using different CCF extraction methods.
\item HAT-P-57\,A is a star with exceptionally high rotational broadening, which makes it an ideal target for testing how the line width of the template spectrum used in cross-correlation affects the resulting CCFs and the visibility of the Doppler shadow.
\item XO-3\,A was selected due to the discrepancy in previously published values of the $\lambda$, for which we wanted to check whether one of the reported values holds up when applying our tomographic method.
\end{enumerate}
\par
Elemental parameters for each system are given in Table \ref{tab:objects_parameters}.
  
\begin{table*}[]
    \caption{Parameters for each system as given in: \textsuperscript{(1)}\cite{Bieryla2014}, \textsuperscript{(2)}\cite{Bourrier2023}, \textsuperscript{(3)}\cite{Hartman2015}, \textsuperscript{(4)}\cite{Winn2009}, \textsuperscript{(5)}\cite{Wong2014}, \textsuperscript{(6)}\cite{Southworth2010}, \textsuperscript{(7)}\cite{Bonomo2017}. Horizontal line divides parameters associated with the planet (upper section) and parameters associated with the star (lower section). All fixed values used in the fitting were taken from this table.}
    \begin{center}
        \begin{tabular}{lccc}
        Parameter & HAT-P-49 & HAT-P-57\,A & XO-3\,A\\
        \hline \hline
        \(P\) (d) & $2.6916\pm6\times10^{-6}$ \textsuperscript{(1)} & $2.4653\pm3.2\times10^{-7}$ \textsuperscript{(3)} & $3.1915\pm6.8\times10^{-8}$ \textsuperscript{(4)} \\ 
        \(T_0\) (BJD) & $	2456399.62406\pm0.00063$ \textsuperscript{(1)} & $2455113.48127\pm0.00048$ \textsuperscript{(3)} & $2454449.86816\pm0.00023$ \textsuperscript{(4)}\\
        \(T_{14}\) (h) & $4.1088\pm0.0456$ \textsuperscript{(1)} & $3.4987\pm0.0192$ \textsuperscript{(3)} & $2.988\pm0.0288$ \textsuperscript{(4)} \\
        \({R_p}/{R_\star}\) & $0.0792\pm0.0019$ \textsuperscript{(1)} & $0.0968\pm0.00015$ \textsuperscript{(3)} & $0.08825\pm0.00037$ \textsuperscript{(5)}\\
        \({a}/{R_\star}\) & $5.13_{-0.30}^{+0.19}$ \textsuperscript{(1)} & $5.825_{-0.116}^{+0.069}$ \textsuperscript{(3)} & $7.052_{-0.097}^{+0.076}$ \textsuperscript{(5)}\\
        \(M\) (M\textsubscript{Jup}) & $1.730\pm0.205$ \textsuperscript{(1)} & $<1.85$ \textsuperscript{(3)} & $11.83\pm0.39$ \textsuperscript{(6)} \\
        \(i\) (deg) & $86.2\pm1.7$ \textsuperscript{(1)} & $88.26\pm0.85$ \textsuperscript{(3)} & $84.20\pm0.54$ \textsuperscript{(4)}\\
        \(e\) & $0$ \textsuperscript{(1)} & $0$ \textsuperscript{(3)} & $0.2769_{-0.0016}^{+0.0017}$ \textsuperscript{(5)}\\
        \(\omega\) (deg) & $90$ \textsuperscript{(2)} & - & $347.2_{-1.6}^{+1.7}$ \textsuperscript{(5)}\\
        \(K\) (m/s) & $188.76\pm21.9$ \textsuperscript{(1)} & $<215.2$ \textsuperscript{(3)} & $1488.0\pm10$ \textsuperscript{(6)}\\
        \(\lambda\) (deg) & $-97.7\pm1.8$ \textsuperscript{(2)} & - & $37.3\pm3.7$ \textsuperscript{(4)}\\
        \hline
        \(v\sin{i_*}\) (km/s) & $16.0\pm0.5$ \textsuperscript{(1)} & $102.1\pm1.3$ \textsuperscript{(3)} & $18.54\pm0.17$ \textsuperscript{(4)}\\
        \(M\) (M\textsubscript{Sun}) & $1.543\pm0.051$ \textsuperscript{(1)} & $1.47\pm0.12$ \textsuperscript{(3)} & $1.213\pm0.066$ \textsuperscript{(4)} \\
        \(R\) (R\textsubscript{Sun}) & $1.833_{-0.076}^{+0.138}$ \textsuperscript{(1)} & $1.50\pm0.05$ \textsuperscript{(3)} & $1.377\pm0.083$ \textsuperscript{(7)} \\
        \(T_{\mathrm{eff}}\) (K) & $6820\pm52$ \textsuperscript{(1)} & $7500\pm250$ \textsuperscript{(3)} & $6429\pm100$ \textsuperscript{(4)} \\ 
        \(Age\) (Gyr) & $1.5\pm0.2$ \textsuperscript{(1)} & $1.00_{-0.51}^{+0.67}$ \textsuperscript{(3)} & $2.82_{-0.82}^{+0.58}$ \textsuperscript{(7)} \\
        \([Fe/H]\) (dex) & $0.074\pm0.080$ \textsuperscript{(1)} & $-0.25\pm0.25$ \textsuperscript{(3)} & $-0.177\pm0.080$ \textsuperscript{(7)} \\
        \end{tabular}
    \end{center}
    \label{tab:objects_parameters}
\end{table*}

\hfill \break
\section{Methodology}
During a planetary transit, the observed line profiles of a star continuously change. Such changes are typically very subtle, so instead of analyzing individual absorption lines directly, the CCF can be used, serving as a proxy for the average stellar line profile. By cross-correlating the observed spectrum with a suitable template, the CCF effectively stacks many spectral lines together, boosting the signal-to-noise ratio and providing a clean, velocity-resolved profile. This averaged profile retains the key broadening and distortion features like stellar pulsations \citep[e.g.][]{Kochukhov2007} or starspots \citep[e.g.][]{KovariZsoltBartus2007}, including — most crucially for our analysis — the partial occultation of the rotating stellar surface by a planetary companion.
\par
In this work, we explored three independent methods for CCF extraction. The first is the HARPS-N data reduction pipeline \citep{Pepe2002}, available on the \texttt{Yabi} platform \citep{YABI} at the IA2 data center\footnote{\url{https://www.ia2.inaf.it/}}. The \texttt{Yabi} pipeline performs cross-correlation using a weighted binary stellar mask specifically tailored to a given spectral type. The cross-correlation is done separately on each spectral order, and a master CCF is then constructed by co-adding the individual-order CCFs. In addition to the \texttt{Yabi} pipeline, we employed two other freely available software tools capable of cross-correlating spectra: \texttt{iSpec} \citep{BCuaresma2014, BCuaresma2019}) which provides Python-based commands, and \texttt{IRAF} \citep{IRAF} using the \texttt{fxcor} task. For both of them, we employed the stitched (1D) spectra as input. Whether synthetic or observational in origin, the cross-correlation templates used in both \texttt{iSpec} and \texttt{IRAF} share the characteristic of containing well-defined, Gaussian-like line profiles.
\par
We visualize the transit as a tightly spaced time series of residual line profiles, obtained by subtracting an average out-of-transit profile from each individual observation, with all profiles placed on a common velocity grid and normalized to a consistent continuum level. This approach highlights the distortion introduced by the transiting planet, which is otherwise absent. To separate the in-transit and out-of-transit data, we determine the phase corresponding to ingress and egress as \(\pm\frac{T_{14}/2}{P/24}\), where $T_{14}$ is the full transit duration in hours and $P$ is the orbital period in days.

\hfill \break
\section{Data Analysis}
\subsection{HAT-P-49}
The hot Jupiter HAT-P-49\,b was first reported by \cite{Bieryla2014} to revolve around one of the most massive stars with a well-determined mass and radius, at an orbital separation of \(\sim0.044\)~au and with the period of \(\sim2.69\)~days. Later, \cite{Bourrier2023} performed the first Doppler tomography to find the projected spin-orbit angle \(\lambda\) to be $-97.7^{\circ}\pm1.8^{\circ}$, using HARPS-N spectroscopic data.
\par
To explore how different approaches to cross-correlation affect the resulting line profiles and Doppler maps, we first processed this dataset using the CCFs provided by the \texttt{Yabi} pipeline. By default, \texttt{Yabi} offers stellar masks corresponding to M2, K5, and G2 spectral types. However, HAT-P-49 has an effective temperature of approximately \(6820~K\) \citep{Bieryla2014}, consistent with an F2-type star according to the Eric Mamajek's compilation \footnote{\url{https://www.pas.rochester.edu/~emamajek/EEM_dwarf_UBVIJHK_colors_Teff.txt}}. For this, we decided to use an externally obtained F6 mask from \cite{Rainer2013}, which represents the closest available match to our target, and incorporated it into the pipeline. Although we ultimately adopted the CCFs produced with the F6 mask for our analysis, we found that the visibility and structure of the Doppler shadow remained, at least for this target, essentially unchanged when using the G2 mask provided by default.
\par
Second, we employed \texttt{iSpec} for which we developed a custom script to automate the cross-correlation, trimming and extraction of CCFs. In this case, we used a rotationally non-broadened synthetic stellar spectrum as the template, created inside \texttt{iSpec} via the SPECTRUM radiative transfer code \citep{SPECTRUM}, while adopting the spectroscopic stellar parameters listed in Table~\ref{tab:objects_parameters}. Solar abundances were taken from \cite{Grevesse1998}, together with the Gaia-ESO Survey (GES) line list. As the grid of model atmospheres, we chose MARCS GES/APOGEE \citep{Plez2008}. The spectral resolution was set to 115000, which matches the resolution of HARPS-N, and the velocity step was set to $0.25$~km/s. All of the stitched 1D spectra provided by the DRS - sampled with 0.60~km/s per pixel at 500~nm - were rebinned accordingly, as this is the procedure also done in the \texttt{Yabi} pipeline prior to computing the CCFs with a numerical mask.

\begin{figure*}[ht]
    \centering
    \includegraphics[width=\linewidth]{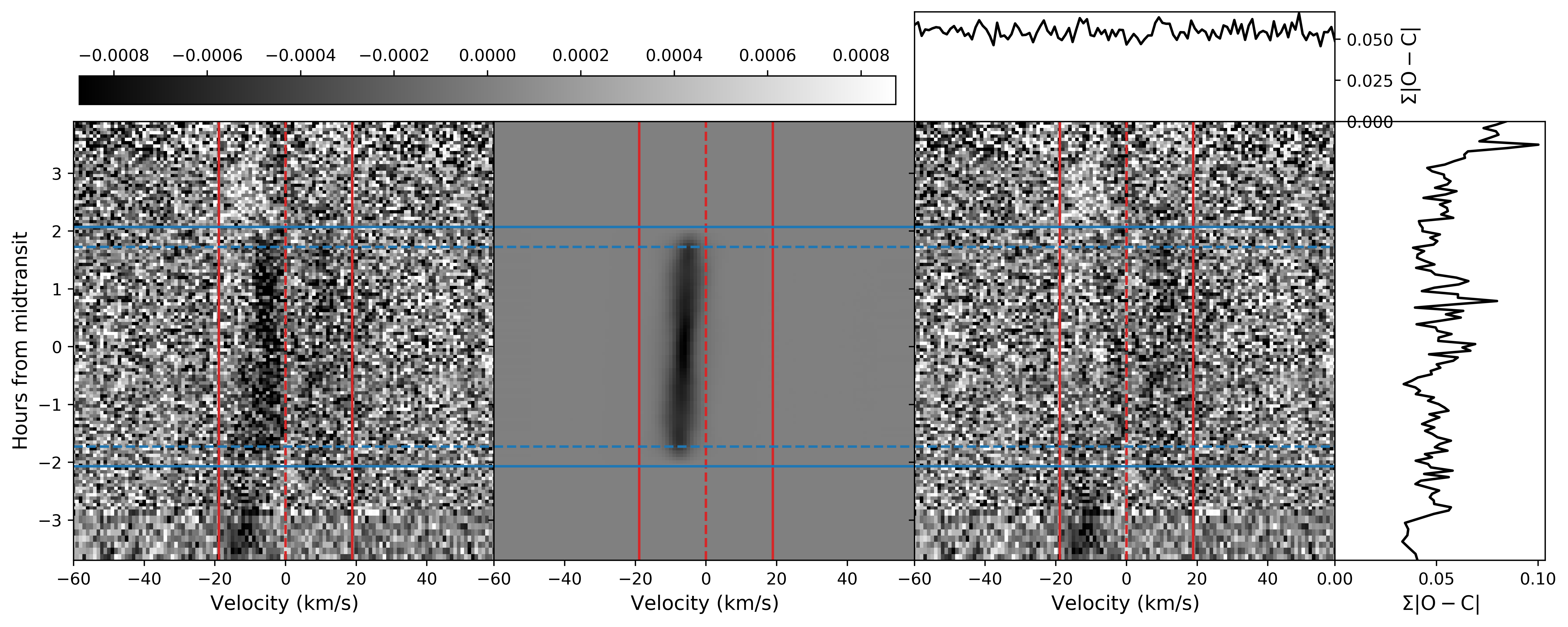}
    \caption{From left to right: data - best-fit model - residuals from fitting the Doppler shadow of HAT-P-49 using the \texttt{Yabi} CCFs on a grid of -60 to +60 km/s. Solid blue horizontal lines depict the first (at the bottom) and last (at the top) planet's contact with the stellar disk, dashed blue lines border the total transit. Solid red lines denote the borders behind which no influence of the planet on the line profiles is observed, i.e., borders for normalization baseline. The zero-velocity represents the center of the stellar disc, and the combination of a high projected obliquity and impact parameter - reflected by inclination - implies that the planet’s trajectory does not intersect the redshifted hemisphere of the star.}
    \label{fig:tracit_fitting_H49}
\end{figure*}

Third, we applied an \texttt{IRAF}-based routine, utilizing the \texttt{fxcor} task. The cross-correlation was carried out using the synthetic spectrum previously constructed in \texttt{iSpec}, with identical binning of the 1D object spectra. Inside both \texttt{iSpec} and \texttt{IRAF}, we restricted the cross-correlation region to 450-580~nm, where the telluric contamination is not severe for neither HAT-P-49, nor any other object analyzed later in this work.

\begin{table*}[]
    \begin{center}
    \caption{Priors and results for fitting the Doppler shadow of HAT-P-49 from three differently obtained sets of CCFs.}
        \begin{tabular}{lcccc}
        Parameter & Prior & Posterior\textsubscript{Yabi CCFs} & Posterior\textsubscript{iSpec CCFs} & Posterior\textsubscript{IRAF CCFs}\\
        \hline \hline
        $v\sin i_*$ (km/s) & $\mathcal{N}(16, 2)$ & $18.9_{-0.8}^{+0.9}$ & $14.5_{-0.8}^{+0.6}$ & $12.2_{-0.6}^{+0.7}$\\
        $\lambda$ (deg) & $\mathcal{U}(-180, 180)$ & $-85.3\pm1.7$ & $-76\pm3$ & $-73_{-4}^{+5}$\\
        $K$ (m/s) & $\mathcal{N}(188.7, 10)$ & $182\pm10$ & $187_{-7}^{+9}$ & $190_{-5}^{+9}$\\
        $\zeta$ (km/s) & $\mathcal{N}(6.0, 1)$ & $7.0\pm0.6$ & $13.8\pm0.5$ & $13\pm0.6$ \\
        $\xi$ (km/s) & $\mathcal{N}(1.5, 0.5)$ & $1.51_{-0.19}^{+0.2}$ & $1.5\pm0.5$ & $1.6_{-0.5}^{+0.3}$ \\
        $q_1+q_2$ & $\mathcal{N}(0.642, 0.1)$ & $0.62_{-0.1}^{+0.09}$ & $0.49\pm0.09$ & $0.41_{-0.08}^{+0.09}$ \\
        \hline
        \end{tabular}
    \label{tab:H49-priors_posteriors}
    \end{center}
\end{table*}

We performed an MCMC analysis implemented in the user-developed Python package \texttt{tracit} (introduced in \cite{Knudstrup2022}), which is suitable for modeling traces of exoplanets (see Figure~\ref{fig:tracit_fitting_H49}). We set 8000 draws and 100 walkers for the sampling method, which was enough to achieve good autocorrelation given the number of fitted parameters. Here, $\lambda$ was set a free parameter. Mean prior values for the star's projected rotational velocity $v\sin i_*$ and the radial velocity semi-amplitude $K$ were taken from Table \ref{tab:objects_parameters}. However, for $v\sin i_*$, we adopted a slightly wider gaussian prior to the mean value reported, to prevent the sampler from being overly constrained by a potentially underestimated uncertainty. In contrast, although \cite{Bieryla2014} reported the uncertainty on $K$ of 21.9~m/s, based on extensive tests of our sampling setup, we found that the posterior constraint on $K$ is consistently much narrower (typically $\sim$10~m/s or less) even when adopting substantially broader priors. This indicates that the data strongly constrain $K$, and therefore we adopted a correspondingly narrower prior without influencing the resulting posterior.
\par
The rest of the fitted parameters are $\zeta$ and $\xi$, denoting the macro- and microturbulences, respectively, and sum of the quadratic limb darkening coefficients, $q_1+q_2$. These were (as will be also for the rest of the analyzed objects) obtained from the ExoCTK tool\footnote{\url{https://exoctk.stsci.edu/limb_darkening}}. All other parameters relevant for fitting were set to the fixed values listed in Table \ref{tab:objects_parameters}. The resulting medians of the posterior distributions and the associated 1$\sigma$ uncertainties of the inferred parameters are contained in Table \ref{tab:H49-priors_posteriors}. 
\par
\begin{figure}[]
    \centering
    \includegraphics[scale = 0.45]{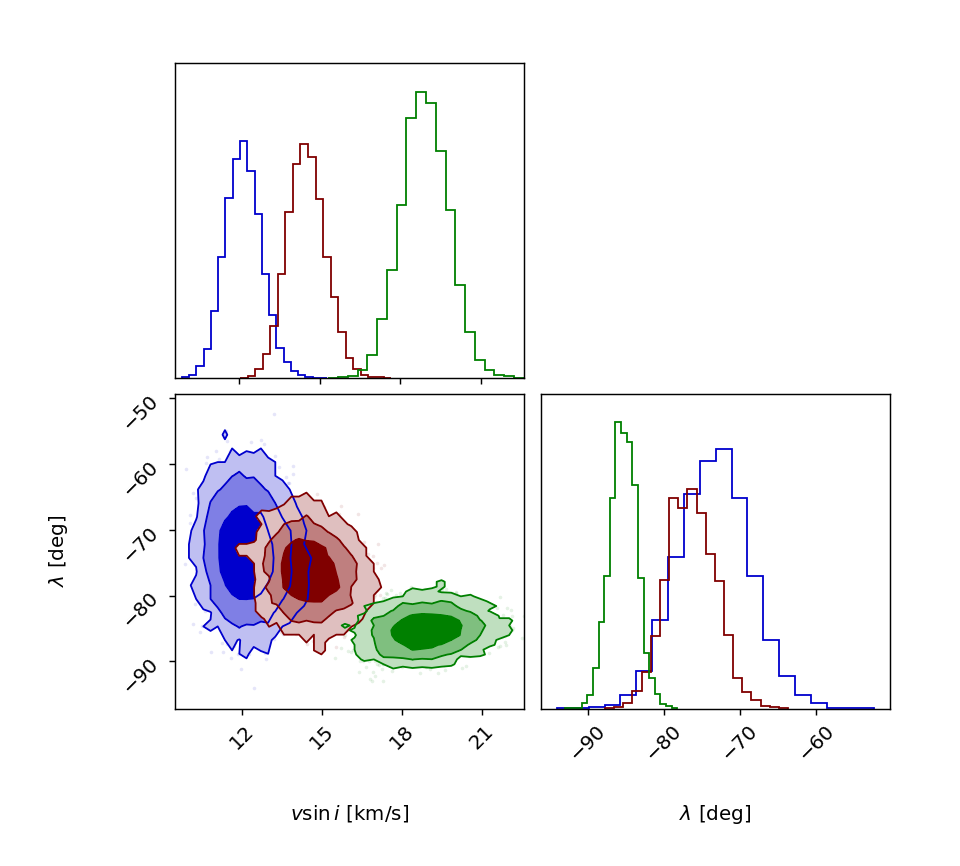}
    \caption{Comparison of fitting three different CCF datasets made out of the same spectral dataset of HAT-P-49; green: \texttt{Yabi} pipeline, blue: \texttt{IRAF}, red: \texttt{iSpec}. Contours denote 1, 2 and 3\,$\sigma$ confidences.}
    \label{fig:H49-lambda_vsini_comparison}
\end{figure}

Despite using the same spectral data as in \cite{Bourrier2023}, not only did we not recover a lambda value approaching their $-97.7,^\circ$ (see Table \ref{tab:objects_parameters}), we also encountered discrepancies between different analyzes. For those, the comparative $v\sin i_*$-$\lambda$ corner plot is depicted in Figure \ref{fig:H49-lambda_vsini_comparison}. Noticeable here is the discrepancy of the $v\sin i_*$ values among different data extraction methods - an effect that has been observed before by \cite{Brown2017} - which decreases at the expense of increasing the velocity of macroturbulences $\zeta$. We also note that the macroturbulence value may appear somewhat elevated, which, however, does not contradict the analysis of \citet{Doyle2014}, who has inferred $\zeta$ from asteroseismic analyses of a sample of F- and G-type stars, demonstrating its positive correlation with effective temperature. We also report that for these analyses, setting uniform prior for both $v\sin i_*$ and $\lambda$ had negligible impact on the final parameter values, compared to the uniform $\lambda$ and gaussian $v\sin i_*$ priors.

To assess the plausibility of the inferred results, we ran additional attempts with a Gaussian prior of $-97.7,^\circ$ set on the spin–orbit angle. For the \texttt{Yabi} CCFs, this fit converged to $-85.9\pm1.5^\circ$, closely matching the result obtained using the uniform prior in Table \ref{tab:H49-priors_posteriors}. In contrast, the \texttt{iSpec} and \texttt{IRAF} CCFs showed a stronger sensitivity to the choice of prior, yielding $-81\pm6,^\circ$ and $-79\pm5,^\circ$, respectively. Along with this, we report that the results obtained using a uniform and a Gaussian prior for \texttt{iSpec} CCFs differ by $0.8-1.7~\sigma$, depending on the adopted uncertainty. For the \texttt{IRAF} CCFs, the corresponding difference is $1.2-1.5~\sigma$. In terms of differences between specific inferred parameters in \texttt{iSpec} and \texttt{IRAF} CCF residuals, the results show visible similarity. However, the extraction process associated with \texttt{fxcor} is considerably more complex and harder to reproduce, for which we prefer to exclude \texttt{IRAF} from the remaining analyses, and instead rely on modern, more user-friendly and well-documented tools. At the same time, we determine the projected spin–orbit angle for HAT-P-49 to be $\lambda = -85.3 \pm 1.7^\circ$, based on the \texttt{Yabi} CCFs fit (see \ref{fig:best_corner_h49}), which is the only case where the result does not appear to be influenced by prior bias.

\hfill \break
\subsection{HAT-P-57\,A}
This system was discovered by \cite{Hartman2015} and reported to host a hot giant planetary companion with an orbital period of $\sim$2.47 days. As they stated at the time of their discovery, it was the fourth exoplanet - found via the transit technique - known to orbit around an A type star. According to the Catalogue of Exoplanets\footnote{\url{https://exoplanet.eu/catalog/}}, the number has increased to 28 as of August 2025.

\begin{figure*}[t]
    \centering
    \includegraphics[width=\linewidth]{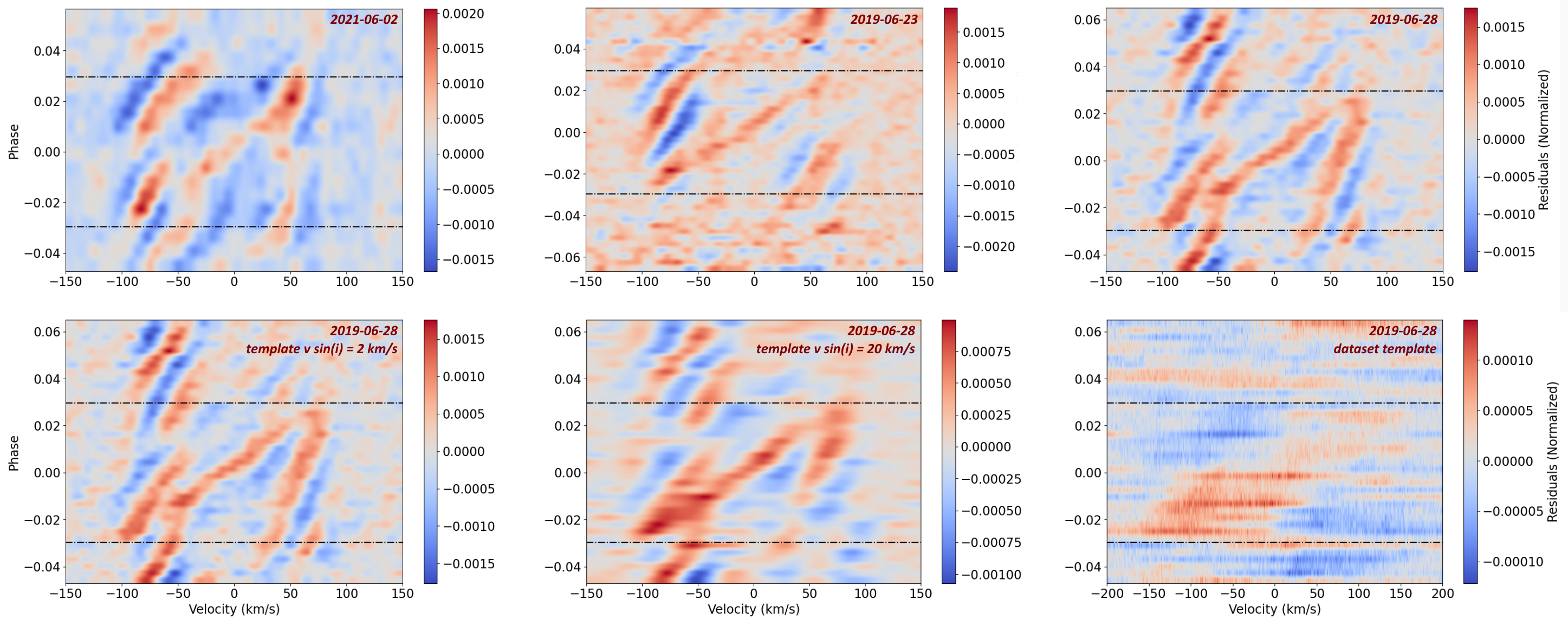}
    \caption{Top row: Doppler tomography of HAT-P-57\,A from three observing nights created from CCFs on sharp-lined synthetic template ($v\sin i_*=2$~km/s). The dashed horizontal lines mark the orbital phases of the first and fourth transit contacts. Bottom row: Third of the selected nights (2019-06-28) plotted as created from CCFs on synthetic templates with $v\sin i_*$ of 2 and 20 km/s, and a dataset template for the last case. Besides the Doppler shadow, we can see additional diagonal distortions with different slope, which extend beyond the transit boundaries. These can be attributed to stellar oscillations.}
    \label{fig:H57-nights_and_templates}
\end{figure*}

In their analysis, \cite{Hartman2015} also modeled the projected spin-orbit angle on the average spectral absorption line profile of a set of nine consecutive Keck-I/HIRES observations, presenting this angle to lay within the intervals of $\langle -16.7\,^\circ < \lambda < 3.3\,^\circ \rangle$ or $\langle 27.6\,^\circ < \lambda < 57.4\,^\circ \rangle$ with 95\,\% confidence. To offer additional insight to this investigation, we conducted our own analysis using available HARPS-N data stored in the TNG archive. By default, these CCF files are produced on a velocity range of $\langle-20, +20\rangle$~km/s. Given the high projected rotational velocity of HAT-P-57\,A ($v\sin i_*\sim102.1$~km/s; \citealt{Hartman2015}), we recalculated the CCFs using the \texttt{Yabi} interface over an expanded interval of 300~km/s centered on the star's systemic velocity. Unfortunately, this procedure eventually did not yield expected results in the form of a planet visibly "traveling" along, but instead becoming indistinguishable within the noisy CCFs. This problem was present both with F6 and G2 stellar masks. Whether - or to what extent - is this issue connected with the stellar type, line broadening, or any other specification, remains unclear. A re-extraction using the new HARPS data-reduction software may provide improved results (or at least explain why the old DRS fails), but this will only be possible in the future, when the updated pipeline becomes accessible via \texttt{Yabi}.
\par
Second approach to obtaining good CCFs was again by \texttt{iSpec}. In order to find out to what extent can the spectral line width in a cross-correlation template affect the detection of the possible shadow, we created (using previously mentioned routines available inside \texttt{iSpec}) a set of synthetic spectra representing a HAT-P-57\,A-like star, but with $v\sin i_*$ of 2 and 20 km/s (bottom row of Figure \ref{fig:H57-nights_and_templates}). Finally, we also synthesized a spectrum with the rotational broadening identical to that of the real star, although, for this case, we ultimately decided to employ an empirical template - to which we will refer as the dataset template - created by averaging five observed out-of-transit spectra with the highest S/N instead. This choice was motivated by the fact that, in the second half of the transit event in our data, the residual map displayed a slightly higher contrast against the background when using the dataset template. However, we recognize that this is not a universal outcome when comparing empirical and synthetic templates, and our preference here reflects the behavior of this particular dataset.
\par
We used data from three all-night observations that occurred on 2021-06-02, 2019-06-23, and 2019-06-28 (upper row in Figure \ref{fig:H57-nights_and_templates}, respectively), and fitted the Doppler shadow using a consistent fitting procedure for all cases. A common feature of each fit was the adoption of a uniform prior distribution for the projected spin–orbit angle $\lambda$. Then, three main fitting configurations were explored, each serving a different purpose.
\par
First, we fitted the shadow on tomography maps constructed using the sharpest available template (i.e., an unbroadened synthetic spectrum). Here we applied a Gaussian prior on the projected stellar rotational velocity $v\sin i_*$, centered at 102.1~km/s (a value reported by \cite{Hartman2015}), elevating the uncertainties slightly more than in the previous object due to the substantially larger rotational broadening and the possible systematics. For the second and third transit nights, this approach yielded similar results for both $\lambda$ and $v\sin i_*$ (top section in Table \ref{tab:H57-priors_posteriors_transposed}). While the $v\sin i_*$ estimated for the first night was also in agreement with the other nights, the inferred $\lambda$ differed by more than 20 degrees, likely due to the planet’s Doppler shadow overlapping with the pulsation features (see Figure \ref{fig:H57-nights_and_templates}, upper left panel). Because the pulsation pattern has a steeper slope than the planetary trail, this overlap may have biased the fit toward a higher absolute value of $\lambda$. Notably, in all three cases, the posterior distributions for $v\sin i_*$ peaked at values significantly lower than the prior mean, not affected by the shape of the prior distribution. When we repeated the analyses with the projected rotational velocity fixed at 102.1~km/s, the inferred $\lambda$ angles for the individual nights were $-34.4\,^\circ$, $19.2\,^\circ$ and $8.5\,^\circ$ (listed in corresponding order in the upper part of Table \ref{tab:H57-priors_posteriors_transposed}), therefore altering the initial results significantly, mainly in the first two cases. 

\begin{figure*}[t]
    \centering
    \includegraphics[width=\linewidth]{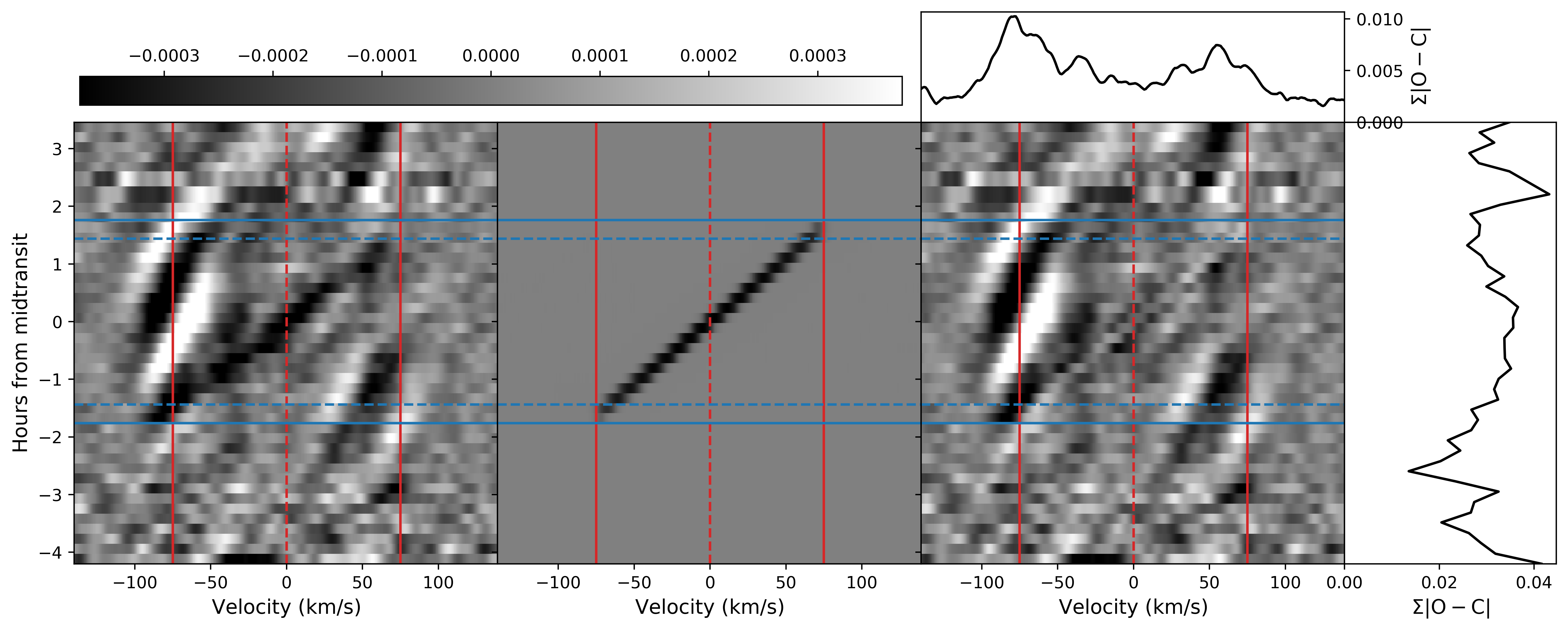}
    \caption{From left to right: data - best-fit model - residuals from fitting the Doppler shadow of HAT-P-57\,A, depicted for 2\textsuperscript{nd} night on a grid of -140 to +140 km/s.}
    \label{fig:tracit_fitting_H57}
\end{figure*}

We also note that the modeling showed itself being largely insensitive to the choice of semi-amplitude $K$. Irrespective of the adopted prior width, the posterior consistently settled near the prior mean. We verified this explicitly by experimenting with substantially broader priors, which resulted in posteriors with comparably large uncertainties, but with only slightly changed central values. The opposite can be said about turbulences, in particular the macroturbulences $\zeta$; because there is no straightforward formula for estimating $\zeta$ in radiative stars, we left this parameter free prior to the MCMC. However, the sampler was unable to constrain it and consistently pushed it toward large values regardless of the allowed range. For that, we adopted a gaussian prior - analogously to the approach used for HAT-P-49. While in that case the posterior of $\zeta$ remained meaningful, we note that in HAT-P-57\,A  this gaussian prior might have been more influential. In addition to these results, we spotted a noticeable deviation of the limb darkening value from the initial estimation, however, we also report that fixing the $q_1+q_2$ did not make any evident impact on the rest of the parameters in other fitting runs.
\par
Sometimes, determining the most reliable best-fit model can be non-trivial. In this case, although the third-night data exhibit a slightly higher average SNR and the Doppler shadow appears with the greatest contrast against the background, it also overlaps with the pulsation patterns visible at the beginning and end of the transit event (see Figure \ref{fig:H57-nights_and_templates}, upper right panel). Therefore, we ultimately selected the fit from the second night as our final result (see Figure \ref{fig:tracit_fitting_H57}), where both the in-transit and out-of-transit line profiles show the least contamination by additional distortions. Moreover, the posterior distributions resulting from this fit appear the most regular and least distorted, further supporting its reliability (see \ref{fig:best_corner_h57}). We therefore adopt a final projected spin-orbit angle of $\lambda = -0.4_{-1.9}^{+1.4}\,^\circ$.
\par
In the second fitting configuration, we used the third-night data to investigate how the use of the rotationally broadened stellar spectrum ($v\sin i_*=20$~km/s) and of the dataset template influences the inferred spin-orbit angle (bottom section in Table \ref{tab:H57-priors_posteriors_transposed}). Compared to the result obtained with the sharpest template ($3.4\pm1.4\,^\circ$), using the synthetic spectrum broadened to $v\sin i_*=20$~km/s resulted in $\lambda=9^{+4}_{-6}\,^\circ$, and using the empirically derived dataset template yielded $\lambda=-7^{+5}_{-10}\,^\circ$. Although — as expected — uncertainties increased with broader templates, all values remained consistent with a well-aligned orbit. Interestingly, the result from the dataset template is in good agreement with the work of \cite{Hartman2015}, who placed $\lambda=-7.7\,^\circ$ among the three highest posterior probabilities (the other two  being $37.6\,^\circ$ and $51.5\,^\circ$). In this fit, however, the $v\sin i_*$ reached as high as 117~km/s, once again not converging to the desired central value. Similar "oscillation" of $\lambda$ and its uncertainties around the initially inferred value, accompanied by an increase in $v\sin i_*$, was observed also in the second-night data, although with a larger deviation, which we attribute to the lower SNR.

\begin{table*}[]
    \begin{center}
    \caption{Priors and results for fitting the Doppler shadow of HAT-P-57\,A; sections are divided by horizontal lines. Top section: Posterior of fitting the tomography on CCFs created using a synthetic template with $v\sin i_*$ of 2~km/s. Bottom section contains - in comparison to the last row in top section - fitting of tomography created on the third-night CCFs made with template with $v\sin i_*$ of 20~km/s, and with averaged dataset template. These two sections essentially represent the fitting of the Doppler shadow for each of the tomographic maps in Figure \ref{fig:H57-nights_and_templates} in corresponding order.} 
    \begin{tabular}{lccccccc}
        Parameter & $v\sin i_*$ (km/s) & $\lambda$ (deg) & $K$ (m/s) & $\zeta$ (km/s) & $\xi$ (km/s) & $q_1 + q_2$ \\
        Prior & $\mathcal{N}(102.1, 5)$ & $\mathcal{U}(-180, 180)$ & $\mathcal{N}(215.2, 10)$ & $\mathcal{N}(6.0, 1)$ & $\mathcal{N}(1.5, 0.5)$ & $\mathcal{N}(0.604, 0.1)$ \\
        \hline \hline
        Posterior\textsubscript{night 1, 2021-06-02} & $90.2_{-1.6}^{+1.7}$ & $-26.58_{-0.13}^{+2.08}$ & $217\pm10$ & $6.54_{-0.2}^{+0.19}$ & $1.58_{-0.2}^{+0.21}$ & $0.29_{-0.1}^{+0.08}$ \\
        Posterior\textsubscript{night 2, 2019-06-23} & $86.5\pm0.8$ & $-0.4_{-1.9}^{+1.4}$ & $216\pm10$ & $6.91_{-0.18}^{+0.19}$ & $1.6_{-0.21}^{+0.2}$ & $0.2_{-0.08}^{+0.05}$ \\
        Posterior\textsubscript{night 3, 2019-06-28} & $87.5_{-1.5}^{+1.1}$ & $3.4\pm1.4$ & $217\pm10$ & $6.64\pm0.19$ & $1.6_{-0.2}^{+0.21}$ & $0.3\pm0.08$ \\
        \hline 
        Posterior\textsubscript{temp. broad. 20 km/s} & $97\pm3$ & $9_{-6}^{+4}$ & $216_{-9}^{+10}$ & $6.12_{-0.21}^{+0.19}$ & $1.55_{-0.21}^{+0.19}$ & $0.45\pm0.09$ \\
        Posterior\textsubscript{dataset temp.} & $117\pm5$ & $-7_{-10}^{+5}$ & $219_{-9}^{+10}$ & $6.01\pm0.2$ & $1.51_{-0.19}^{+0.21}$ & $0.118_{-0.018}^{+0.011}$ \\
        \hline 
        \end{tabular}
    \label{tab:H57-priors_posteriors_transposed}
    \end{center}
\end{table*}

Finally, we repeated the second fitting setup (that is, we investigated the third-night data constructed with all of the three cross-correlation templates), but forcing the $v\sin i_*$ of the modeled CCFs to be fixed at 102.1~km/s. Not only did this constraint lead to higher values of the spin-orbit angle ($\lambda=11.783_{-0.017}^{+0.027}\,^\circ$ using the synthetic spectrum with $v\sin i_*=2$~km/s, and $\lambda=12.1_{-0.4}^{+3.9}\,^\circ$ using the synthetic spectrum with $v\sin i_*=20$~km/s), but also resulted in posterior distributions characterized by one (for the first case) and two closely spaced (for the second case) exceptionally sharp peaks, which in turn produced unrealistically narrow 1$\sigma$ uncertainties. A similar behavior was observed in the $\lambda$ posteriors obtained from fitting the tomography made with the dataset template, which, however, cannot be considered statistically reliable, as the corresponding reduced $\chi^2$ exceeded one hundred. In comparison, for the previous fits the averaged reduced $\chi^2$ was 1.24. Even doubling the number of draws in this MCMC analysis attempt to 16000 did not bring us closer to meaningful results, for which we do not include it among our primary set of solutions.

\hfill \break
\subsubsection{The Pulsations of HAT-P-57\,A}
In addition to the line-profile variations caused by the planet seen as the central strip in the Doppler tomography (Figs.~\ref{fig:H57-nights_and_templates} and \ref{fig:tracit_fitting_H57}), there is a regular structure apparent to the left but also to the right from the center, although with lower amplitude. Similar structures were interpreted as the manifestation of non-radial pulsations in, for example, HD\,49434 \citep{Uytterhoeven2008} and HD\,15082 \citep{CollierCameron2010B}. The larger amplitudes in the blue part were reported by \citet{Mathias2004} and explained by the equivalent width variations due to pulsations \citep{Schrijvers1999}. Since the pattern shown in Fig.~\ref{fig:H57-nights_and_templates} is stable with respect to the transit during all three nights, we may speculate that the pulsations are linked with the orbital motion of the planet via tidal interactions.

\begin{figure}[t]
    \centering
    \includegraphics[scale = 0.5]{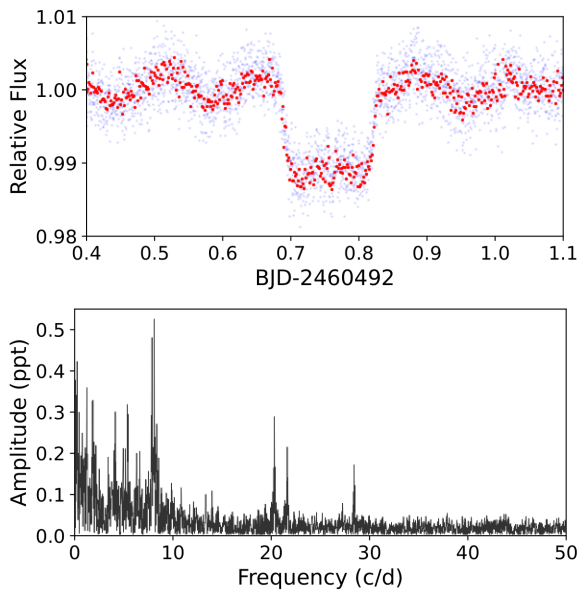}\\
    \caption{Top panel: Sector 80 \textit{TESS} data with 20-seconds (red) and 120-seconds cadences (blue) around one of the observed transits. The frequency spectrum of the 20-seconds data once the transits have been masked out is shown in the bottom panel.}
    \label{Fig:Photometry}
\end{figure}

The temperature 7500\,K of HAT-P-57\,A is close to the peak of the temperature distribution of pulsating $\delta$\,Sct and hybrid $\gamma$\,Dor/$\delta$\,Sct stars \citep[e.g.][]{Murphy2019,Skarka2022}. Thus, stellar oscillations in HAT-P-57\,A are expected and were already predicted by \citet{Hartman2015}. HAT-P-57\,A was observed by \textit{TESS} \citep{Ricker2015} only in Sector 80. Data products with 120- and 20-seconds cadences from the SPOC routine \citep{Jenkins2016} show fast variations easily detectable in the light curve but also in the periodogram (Fig.~\ref{Fig:Photometry}). After we manually masked out transits, the frequency spectrum (bottom panel of Fig.~\ref{Fig:Photometry}) clearly shows multiple peaks with the dominant frequency at $f_{\rm puls}=8.0935$\,c/d that is typical for $\delta$\,Sct pulsations \citep[e.g.][]{Breger2000,Antoci2019}. Interestingly, the dominant pulsation frequency is approximately 20-times the orbital frequency ($f_{\rm puls}\approx 20 f_{\rm orb}$) and frequency at 20.316\,c/d is $\sim50f_{\rm orb}$. This strengthens the assumption that there might be a tidal interaction and link between orbital and pulsations. There are also peaks in the $\gamma$\,Dor regime (below 5\,c/d) providing hints that HAT-P-57\,A is a hybrid pulsator \citep[e.g.][]{Uytterhoeven2011,Antoci2019}. There might be also peaks induced by rotation in the low-frequency regime. However, it is not straightforward to identify them, particularly when the rotation may be tidally locked with the orbital frequency. From the light curve (top panel of Fig.~\ref{Fig:Photometry}), it is obvious that the pulsations significantly affect transits. HAT-P-57\,A system deserves detail analysis of the \textit{TESS} data, pulsations and tidal interaction, that is out of scope of this paper.

\hfill \break
\subsection{XO-3\,A}
After the discovery of the companion XO-3\,A\,b on an orbit of $\sim3.19$ days by \cite{Johns-Krull2008}, a number of other studies followed \citep[e.g.][]{Winn2008, Wong2014, Southworth2010}. While all of these confirmed the object to lie near the planet–brown dwarf boundary ($\approx12$~M\textsubscript{Jup}), \cite{Stassun2017} later reported a significantly lower mass of $\sim7.3~$~M\textsubscript{Jup}. Another interesting discrepancy - the most relevant for our work - emerged among the measurements of the projected spin-orbit angle. The first result presented of $\lambda$ for this system, based on SOPHIE spectrograph observations, was presented by \cite{Hebrard2008}, who derived a value of $70\pm15\,^\circ$ from modeling the RV-RM effect. However, they noted that this result could be biased due to significant variations in airmass experienced during the observation. A subsequent study by \cite{Winn2009}, using Keck/HIRES data, yielded a revised spin-orbit angle of $37\pm3.7\,^\circ$. \cite{Hirano2011} later supported this result, while also pointing out the presence of a radial velocity trend between subsequent epochs, suggesting the possible existence of an additional body in the system. During the preparation of this study, \cite{Rusznak2025} performed the first Doppler tomography analysis of XO-3\,A using NEID spectrograph data, reporting the projected spin-orbit angle of $40_{-2.0}^{+2.1}\,^\circ$, which is in close agreement with previously reported value of $37\pm3.7\,^\circ$.

\begin{figure*}[]
    \centering
    \includegraphics[width=\linewidth]{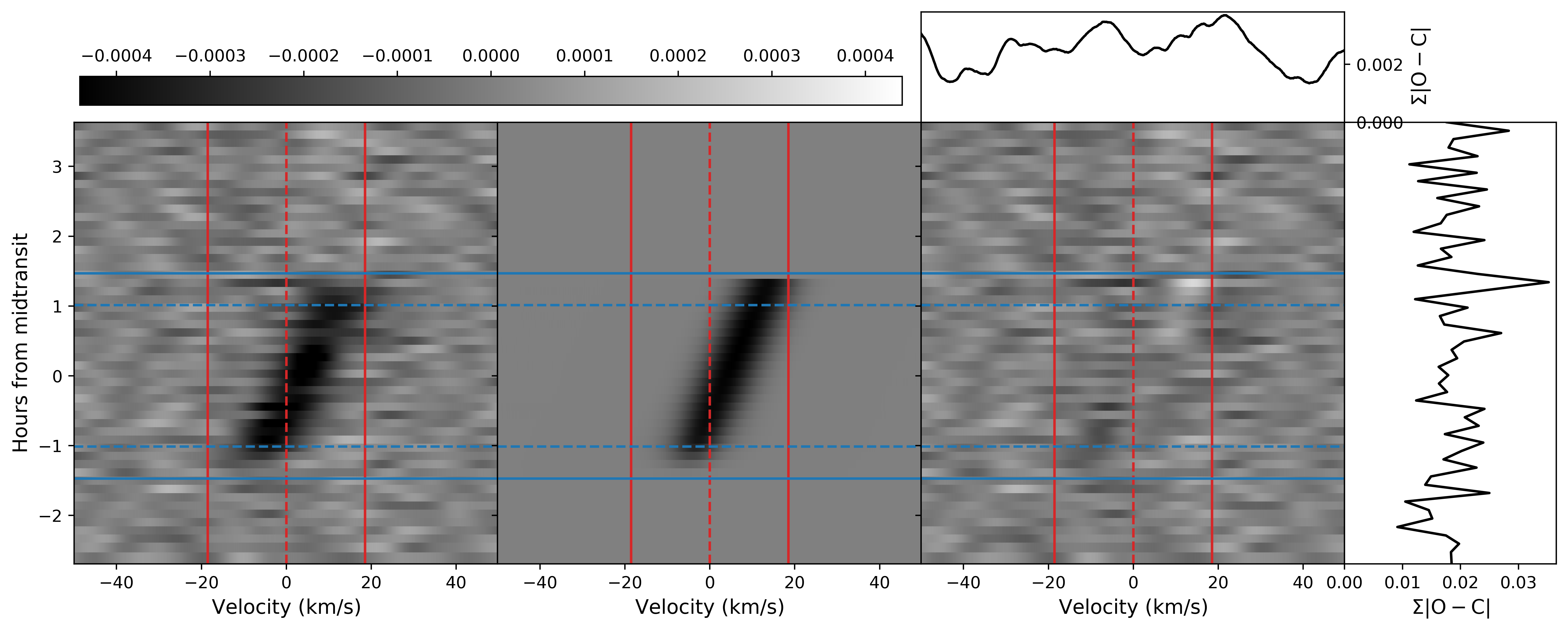}
    \caption{From left to right: data - best-fit model - residuals from fitting the Doppler shadow of XO-3\,A; based on tomography constructed from CCFs produced with \texttt{iSpec}.}
    \label{fig:tracit_fitting_XO3}
\end{figure*}

\begin{table*}[]
    \begin{center}
    \caption{Priors and results for fitting the Doppler shadow of XO-3\,A from two differently obtained sets of CCFs}
        \begin{tabular}{lccc}
        Parameter & Prior & Posterior\textsubscript{Yabi CCFs} & Posterior\textsubscript{iSpec CCFs} \\
        \hline \hline
        $v\sin i_*$ (km/s) & $\mathcal{N}(18.54, 2)$ & $17.2\pm1.1$ & $18.2_{-1.2}^{+1.1}$ \\
        $\lambda$ (deg) & $\mathcal{U}(-180, 180)$ & $43\pm6$ & $38_{-4}^{+3}$ \\
        $K$ (m/s) & $\mathcal{N}(1488, 5)$ & $1487\pm5$ & $1488\pm5$ \\
        $\zeta$ (km/s) & $\mathcal{N}(6.0, 1)$ & $7.4_{-0.7}^{+0.6}$ & $11.0\pm0.6$ \\
        $\xi$ (km/s) & $\mathcal{N}(1.5, 0.5)$ & $2.2\pm1.0$ & $1.7_{-1.0}^{+0.8}$ \\
        $q_1+q_2$ & $\mathcal{N}(0.628, 0.1)$ & $0.61_{-0.09}^{+0.10}$ & $0.65_{-0.11}^{+0.09}$ \\
        \hline
        \end{tabular}
    \label{tab:XO3-priors_posteriors}
    \end{center}
\end{table*}

We used \texttt{Yabi} with the F6 stellar mask obtained from \cite{Rainer2013} to create CCFs on a $\langle-60, +60\rangle$~km/s velocity range, ensuring a wide enough baseline on both sides of the profile. However, due to an unexplainable issue in the pipeline, several of the output data showed a general shift along the x-axis, which produced severe outliers in the tomography map that we could not appropriately fix by additional means. Therefore, we were forced to discard a significant portion of the dataset, ultimately using only 37 out of the original 54 CCFs (see Fig \ref{fig:tracit_fitting_XO3}), including 14 of the 21 in-transit frames. Fortunately, producing the CCFs via \texttt{iSpec} did not yield such outliers, which left us with the full night data. Considering our previous experience, using a template broadened to match the star’s actual line broadening tends to yield $\lambda$ values with the largest uncertainties - therefore, we omit such approach this time.
\par
Utilizing our knowledge about the behavior of the sampler discussed in the previous sections, we used \texttt{tracit} again and performed the MCMC analysis on both of our residual data. To not give prior preference to any of the earlier reported and very different projected spin-orbit angles, we set for this parameter a uniform prior distribution, finishing with 8000 draws, as in the previous cases. The results (see Table \ref{tab:XO3-priors_posteriors}) indicate that the projected spin-orbit angle is inclined by around $40\,^\circ$, again supporting the value derived by \cite{Winn2009} using the traditional method, and in excellent agreement with Doppler tomography applied on the NEID data by \cite{Rusznak2025}. 
\par
Because of lower amount of evenly-spaced data among the \texttt{Yabi} CCFs, we decided to consider the spin-orbit angle fit of the \texttt{iSpec} CCFs as the final result (see \ref{fig:best_corner_xo3}), giving $\lambda=38_{-4}^{+3}\,^\circ$.

\hfill \break
\section{Timescales}
\begin{figure*}
    \centering
    \includegraphics[width=\linewidth]{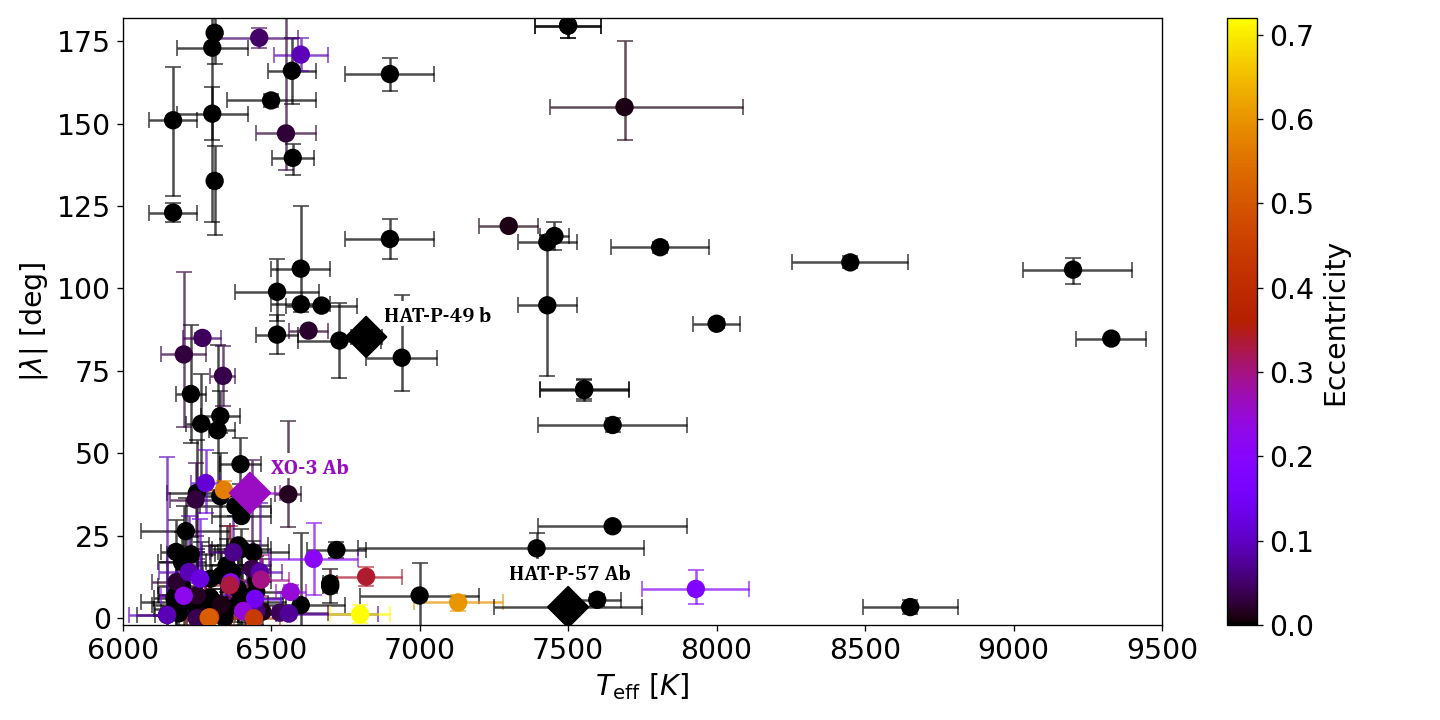}
    \caption{Sky-projected spin-orbit angles $|\lambda|$ as functions of stellar effective temperatures \(T_{\mathrm{eff}}\) for stars above 6250~K. Dots denote the known obliquities listed in the TEPCat catalogue\footnote{\url{https://www.astro.keele.ac.uk/jkt/tepcat/obliquity.html} (8/8/2025)}, diamonds depict the three objects analyzed in this work. Colorful markers belong to planets with reported eccentricity.}
    \label{fig:obliquities}
\end{figure*}
We plot the result from our analysis among other known measurements of the projected spin-orbit angle in Figure \ref{fig:obliquities}. Given the small size of our sample, we do not attempt any statistical interpretation; however, it does illustrate the diversity of spin–orbit configurations among hot Jupiters orbiting stars with predominantly radiative envelopes, ranging from well-aligned to highly misaligned and even polar orbits. We also show the known eccentricities of these planets to highlight the fraction that may have undergone high-eccentricity migration.
\par
To assess what are the plausible evolutionary pathways of our sample of planets, we must first determine the relevant timescales in their history. First, we assume the effect of large oscillations in eccentricity and inclination induced by the Kozai-Lidov mechanism in multiple-star systems \citep{Fabrycky2007} - in our case, this applies to the HAT-P-57\,A and XO-3\,A systems. The timescale of these cycles, as given in \cite{Rice2023} (first introduced by \cite{Naoz2016}), can be calculated as
\begin{equation}
\tau_{\mathrm{KL}} \approx \frac{16}{30\pi} \frac{P_B^2}{P_b} \left(1 - e_B^2\right)^{3/2} \left( \frac{M_A + M_B}{M_B} \right).
\end{equation}

We first apply this expression to HAT-P-57\,A. In their analysis, \cite{Hartman2015} found two additional stellar companions of this star at the distance of $\sim$800~au and period $P$ of $\sim$14000 years. Since their mutual distance is very small (only 68~au), we approximate them as one body denoted 'B', with mass $M_\mathrm{B}=0.61+0.53$~M\textsubscript{Sun} (corresponding to the sum of the stars' individual masses), and continue by substituting the main star's mass $M_A$ and the planet's period $P_b$ (see Table \ref{tab:objects_parameters}), along with period $P_B$. Since we do not know anything about the eccentricity $e_B$, we take into account two extreme options, eccentricities of 0 and 0.9, which leads us to a Kozai-Lidov timescale of $\sim1$ - 11~Gyr. This suggests that, in the case of a star whose age approaches the lower boundary of the estimated $\tau_{\mathrm{KL}}$ interval \citep{Hartman2015}, the mechanism has likely not had enough time to significantly affect the system’s evolution.
\par
\cite{Rusznak2025} also mentions a faint M4-M5 companion of XO-3\,A at the distance of $\sim90$~au. Substituting an estimated mass of 0.3~M\textsubscript{Sun} for this star — an approximation based on its spectral type — and applying Kepler's third law, we retrieve an orbital period $P_B$ of 690 years. Using the same range of assumed eccentricities as for HAT-P-57\,A, we then estimate the timescale of a Kozai–Lidov cycle for this system to be $\sim4$ - 45~Myr. Considering XO-3\,A’s older age of 2.8~Gyr \citep{Bonomo2017}, and the planet's high spin-orbit angle, the influence of this mechanism on the system's dynamical evolution appears quite realistic. Furthermore, \cite{Hwang2020} reported another companion of XO-3\,A, a late F-type star that lies 38.000~au away. It is possible that this stellar  configuration could cause an "eccentricity cascade" mechanism, proposed by \cite{Yang2025}, in which the outer star periodically excites the eccentricity of the inner companion, and this eccentricity excitation is slowly transferred to the cold Jupiter, triggering its HEM. If this was the case, it would be the second  system of such type.
 
\par
Next, we determine the timescale of orbital circularization. From our selection of objects, XO-3\,A\,b is the only planet with reported - and quite significant - orbital eccentricity $0.2769_{-0.0016}^{+0.0017}$ (see Table \ref{tab:objects_parameters}). According to
\begin{equation}
    \tau_{\text{circ}} \sim \frac{e_b}{\frac{de_b}{dt}},
\end{equation}

which is a formula described in detail in \cite{Rice2023}, we calculate the circularization time $\tau_{\text{circ}}$ for the planetary orbit to take $\sim22$~Myr - 2.2~Gyr. The uncertainty of the $10^2$ factor arises from estimations of planetary tidal response parameters that are also discussed in \cite{Rice2023} and \cite{Penev2018}. Theoretically, if HAT-P-49\,b and HAT-P-57\,A\,b, had the same eccentricity, their orbits would circularize in $\sim0.8$ - 80~Myr and $\sim0.6$ - 60~Myr, respectively. Relative to the ages of these systems, if such eccentricity was ever real, this dynamical process might have been finished a long time ago.
\par
Finally, the timescale for the alignment of radiative stars (that is, stars above the Kraft break) can be predicted following the expression
\begin{equation}
    \frac{1}{\tau_{\mathrm{RA}}} = \frac{1}{0.25 \cdot 5 \cdot 10^9\,\mathrm{yr}} \, q^2 (1+q)^{5/6} \left( \frac{a / R_\star}{6} \right)^{-17/2},
\end{equation}

given in \cite{Albrecht2012}, where $q\equiv{M_p}/{M_\star}$. We retrieve the values of $2.9\times10^5$~Gyr, $6.8\times10^5$~Gyr and $5.7\times10^4$~Gyr years for HAT-P-49, HAT-P-57\,A and XO-3\,A, respectively. Since these timescales surpass the current age of the universe, the realignment is unlikely to have played a relevant role in shaping the observed configuration of any of these systems. For comparison, a similar calculation applied to a convective star - where tidal dissipation is more efficient — could result in a timescale up to four orders of magnitude shorter \citep{Albrecht2012}.

\hfill \break
\section{Discussion}
Our Doppler tomographic analysis of HAT-P-49, HAT-P-57\,A, and XO-3\,A provides insight into both the orbital architectures of these hot Jupiter systems and the influence of methodological choices on spin–orbit angle measurements. The first of the intriguing examples is the HAT-P-49 system, where we tested three independent routines for CCF extraction: \texttt{Yabi} (i.e., pipeline for HARPS-N used with a G2/F6 weighted binary stellar mask), \texttt{iSpec} and \texttt{IRAF} (both used with a synthetic template representing a similar star). The derived values of $\lambda$ from the respective tomography maps, set with a free prior on this key parameter, vary only by several degrees and are all consistent with a highly oblique, polar orbit. Additional tests with a Gaussian prior on $\lambda$ revealed that $\lambda_{\rm Yabi}$ converged to a similar value in this setup as well, whereas $\lambda_{\rm iSpec}$ and $\lambda_{\rm IRAF}$ shifted more noticeably. This stability makes the $\lambda_{\rm Yabi}$ value more reliable, and we therefore adopt $\lambda_{\rm Yabi}=-85.3\pm1.7^\circ$ as the result most representative of the model. We also conclude that the planet’s high $\lambda$, together with no reported eccentricity and/or stellar companion, are compatible with either a primordial polar origin \citep{Kuffmeier2024} or past HEM with subsequent circularization.
\par
In the case of HAT-P-57\,A, we used a single CCF extraction routine, \texttt{iSpec}, but applied differently broadened template spectra (two synthetic with $v\sin i_*$ of 2 and 20~km/s and one co-added averaged spectrum from the dataset) across three separate transit nights. The use of sharp-lined versus broadened templates did not significantly alter the qualitative spin–orbit classification, as all the resulting fits consistently pointed toward an orbit aligned with the star's rotation. However, it demonstrated that even with consistent processing, template selection can indeed introduce subtle variations and raise uncertainties in $\lambda$ measurements. From one of the CCFs datasets created with the sharpest line template, we chose the best-fit value of $\lambda = -0.4_{-1.9}^{+1.4}\,^\circ$. Such low projected spin-orbit angle implies either formation of the planet in close vicinity to the host star, or migration within the aligned protoplanetary disk.
\par
Finally, XO-3\,A, has been the subject of previous debate, with two significantly different values of $\lambda$ reported in the literature, one near $37\,^\circ$ and another near $70\,^\circ$, leading to conflicting interpretations of the system's spin–orbit architecture. By applying two distinct CCF extraction methods (via \texttt{Yabi} and \texttt{iSpec}) and fitting the Doppler shadow directly, we recovered $\lambda=38_{-4}^{+3}\,^\circ$, representing a moderately misaligned orbit supporting the first measurement mentioned. In this case, this misalignment points to an active Kozai–Lidov mechanism induced by a nearby companion, hinting a disc-free migration.
\par
Knowledge of projected spin-orbit angle $\lambda$ can in some cases lead to determining the true spin-orbit angle $\psi$. However, this requires determining the star's inclination towards the observer - $i_*$ - which is not a trivial task to accomplish, as it depends on detectable photometric variability caused by starspots. None of the star in our sample shows a clear signs of rotational variability that would allow us to measure the rotational period from photometry. Therefore, all of our estimated spin-orbit angles stay in the projected plane.
\par
A complementary approach to study the migration pathways is through the composition of planetary atmospheres by measuring the atmospheric water abundance \citep{cow25} or the elemental abundance ratios such as C/O, N/O and others \citep{turr21}. The BOWIE-ALIGN survey \citep{kirk24} is investigating the atmospheric composition to stellar obliquity link at great detail on sample of 8 systems. The upcoming \textit{Ariel} mission \citep{tin18} is expected to observe up to 1000 of exoplanets in a homogeneous way to study their atmospheric composition and study where and how these planet have formed. All three targets in this work are listed among Tier 3 (high priority targets selected for detailed characterization) sample of the \textit{Ariel} mission target candidates\footnote{\url{https://github.com/arielmission-space/Mission_Candidate_Sample/tree/main/target_lists}} \citep{edw22}. The majority of the \textit{Ariel} target candidates do not have their spin-orbit angle determined \citep{zak25b}. Hence, the measured values of $\lambda$ in this work will provide a key contextual information of the dynamics of these systems and can enable a more robust interpretation of the atmospheric data.

The approach demonstrated in this work can be extended to a wider range of systems with high-resolution spectroscopic data, including those from ongoing surveys and future instruments. Applying Doppler tomography to a larger and more diverse sample will be essential to refine statistical trends in stellar obliquity and strengthen the link between spin–orbit architecture and planetary migration pathways.

\begin{acknowledgments}
The authors thank the Italian Center for Astronomical Archive (IA2, https://www.ia2.inaf.it/), operated by INAF at the Astronomical Observatory of Trieste, for providing support, access to the Yabi service and the use of their facilities. ZB, PK and JZ would like to acknowledge support from a GACR grant: 22-30516K. The research of PG was supported by the internal grant No. VVGS-2023-2784 of the P. J. {\v S}af{\'a}rik University in Ko{\v s}ice funded by the EU NextGenerationEU through the Recovery and Resilience Plan for Slovakia under the project No. 09I03-03-V05-00008.
\end{acknowledgments}

\bibliography{sample701}{}

@INPROCEEDINGS{Breger2000,
       author = {{Breger}, M.},
        title = "{{\ensuremath{\delta}} Scuti stars (Review)}",
    booktitle = {Delta Scuti and Related Stars},
         year = 2000,
       editor = {{Breger}, Michel and {Montgomery}, Michael},
       series = {Astronomical Society of the Pacific Conference Series},
       volume = {210},
        month = jan,
        pages = {3},
       adsurl = {https://ui.adsabs.harvard.edu/abs/2000ASPC..210....3B}
}

@ARTICLE{Uytterhoeven2011,
       author = {{Uytterhoeven}, K. and {Moya}, A. and {Grigahc{\`e}ne}, A. and {Guzik}, J.~A. and {Guti{\'e}rrez-Soto}, J. and {Smalley}, B. and {Handler}, G. and {Balona}, L.~A. and {Niemczura}, E. and {Fox Machado}, L. and {Benatti}, S. and {Chapellier}, E. and {Tkachenko}, A. and {Szab{\'o}}, R. and {Su{\'a}rez}, J.~C. and {Ripepi}, V. and {Pascual}, J. and {Mathias}, P. and {Mart{\'\i}n-Ru{\'\i}z}, S. and {Lehmann}, H. and {Jackiewicz}, J. and {Hekker}, S. and {Gruberbauer}, M. and {Garc{\'\i}a}, R.~A. and {Dumusque}, X. and {D{\'\i}az-Fraile}, D. and {Bradley}, P. and {Antoci}, V. and {Roth}, M. and {Leroy}, B. and {Murphy}, S.~J. and {De Cat}, P. and {Cuypers}, J. and {Kjeldsen}, H. and {Christensen-Dalsgaard}, J. and {Breger}, M. and {Pigulski}, A. and {Kiss}, L.~L. and {Still}, M. and {Thompson}, S.~E. and {van Cleve}, J.},
        title = "{The Kepler characterization of the variability among A- and F-type stars. I. General overview}",
      journal = {\aap},
         year = 2011,
        month = oct,
       volume = {534},
          eid = {A125},
        pages = {A125},
          doi = {10.1051/0004-6361/201117368},
archivePrefix = {arXiv},
       eprint = {1107.0335},
 primaryClass = {astro-ph.SR},
       adsurl = {https://ui.adsabs.harvard.edu/abs/2011A&A...534A.125U}
}

@ARTICLE{Antoci2019,
       author = {{Antoci}, V. and {Cunha}, M.~S. and {Bowman}, D.~M. and {Murphy}, S.~J. and {Kurtz}, D.~W. and {Bedding}, T.~R. and {Borre}, C.~C. and {Christophe}, S. and {Daszy{\'n}ska-Daszkiewicz}, J. and {Fox-Machado}, L. and {Garc{\'\i}a Hern{\'a}ndez}, A. and {Ghasemi}, H. and {Handberg}, R. and {Hansen}, H. and {Hasanzadeh}, A. and {Houdek}, G. and {Johnston}, C. and {Justesen}, A.~B. and {Kahraman Alicavus}, F. and {Kotysz}, K. and {Latham}, D. and {Matthews}, J.~M. and {M{\o}nster}, J. and {Niemczura}, E. and {Paunzen}, E. and {S{\'a}nchez Arias}, J.~P. and {Pigulski}, A. and {Pepper}, J. and {Richey-Yowell}, T. and {Safari}, H. and {Seager}, S. and {Smalley}, B. and {Shutt}, T. and {S{\'o}dor}, A. and {Su{\'a}rez}, J. -C. and {Tkachenko}, A. and {Wu}, T. and {Zwintz}, K. and {Barcel{\'o} Forteza}, S. and {Brunsden}, E. and {Bogn{\'a}r}, Z. and {Buzasi}, D.~L. and {Chowdhury}, S. and {De Cat}, P. and {Evans}, J.~A. and {Guo}, Z. and {Guzik}, J.~A. and {Jevtic}, N. and {Lampens}, P. and {Lares Martiz}, M. and {Lovekin}, C. and {Li}, G. and {Mirouh}, G.~M. and {Mkrtichian}, D. and {Monteiro}, M.~J.~P.~F.~G. and {Nemec}, J.~M. and {Ouazzani}, R. -M. and {Pascual-Granado}, J. and {Reese}, D.~R. and {Rieutord}, M. and {Rodon}, J.~R. and {Skarka}, M. and {Sowicka}, P. and {Stateva}, I. and {Szab{\'o}}, R. and {Weiss}, W.~W.},
        title = "{The first view of {\ensuremath{\delta}} Scuti and {\ensuremath{\gamma}} Doradus stars with the TESS mission}",
      journal = {\mnras},
         year = 2019,
        month = dec,
       volume = {490},
       number = {3},
        pages = {4040-4059},
          doi = {10.1093/mnras/stz2787},
archivePrefix = {arXiv},
       eprint = {1909.12018},
 primaryClass = {astro-ph.SR},
       adsurl = {https://ui.adsabs.harvard.edu/abs/2019MNRAS.490.4040A}
}

@INPROCEEDINGS{HARPSN,
       author = {{Latham}, David W. and {HARPS-N Collaboration}},
        title = "{HARPS-N: A New Tool for Characterizing Kepler Planets}",
    booktitle = {American Astronomical Society Meeting Abstracts \#221},
         year = 2013,
       series = {American Astronomical Society Meeting Abstracts},
       volume = {221},
        month = jan,
          eid = {231.02},
        pages = {231.02},
       adsurl = {https://ui.adsabs.harvard.edu/abs/2013AAS...22123102L}
}

@INPROCEEDINGS{HARPSN_Cosentino,
       author = {{Cosentino}, Rosario and {Lovis}, Christophe and {Pepe}, Francesco and {Collier Cameron}, Andrew and {Latham}, David W. and {Molinari}, Emilio and {Udry}, Stephane and {Bezawada}, Naidu and {Black}, Martin and {Born}, Andy and {Buchschacher}, Nicolas and {Charbonneau}, Dave and {Figueira}, Pedro and {Fleury}, Michel and {Galli}, Alberto and {Gallie}, Angus and {Gao}, Xiaofeng and {Ghedina}, Adriano and {Gonzalez}, Carlos and {Gonzalez}, Manuel and {Guerra}, Jose and {Henry}, David and {Horne}, Keith and {Hughes}, Ian and {Kelly}, Dennis and {Lodi}, Marcello and {Lunney}, David and {Maire}, Charles and {Mayor}, Michel and {Micela}, Giusi and {Ordway}, Mark P. and {Peacock}, John and {Phillips}, David and {Piotto}, Giampaolo and {Pollacco}, Don and {Queloz}, Didier and {Rice}, Ken and {Riverol}, Carlos and {Riverol}, Luis and {San Juan}, Jose and {Sasselov}, Dimitar and {Segransan}, Damien and {Sozzetti}, Alessandro and {Sosnowska}, Danuta and {Stobie}, Brian and {Szentgyorgyi}, Andrew and {Vick}, Andy and {Weber}, Luc},
        title = "{Harps-N: the new planet hunter at TNG}",
    booktitle = {Ground-based and Airborne Instrumentation for Astronomy IV},
         year = 2012,
       editor = {{McLean}, Ian S. and {Ramsay}, Suzanne K. and {Takami}, Hideki},
       series = {Society of Photo-Optical Instrumentation Engineers (SPIE) Conference Series},
       volume = {8446},
        month = sep,
          eid = {84461V},
        pages = {84461V},
          doi = {10.1117/12.925738},
       adsurl = {https://ui.adsabs.harvard.edu/abs/2012SPIE.8446E..1VC}
}

@ARTICLE{attia23,
       author = {{Attia}, M. and {Bourrier}, V. and {Delisle}, J. -B. and {Eggenberger}, P.},
        title = "{DREAM: II. The spin{\textendash}orbit angle distribution of close-in exoplanets under the lens of tides}",
      journal = {\aap},
         year = 2023,
        month = jun,
       volume = {674},
          eid = {A120},
        pages = {A120},
          doi = {10.1051/0004-6361/202245237},
archivePrefix = {arXiv},
       eprint = {2305.00829},
 primaryClass = {astro-ph.EP},
       adsurl = {https://ui.adsabs.harvard.edu/abs/2023A&A...674A.120A}
}

@ARTICLE{alb21,
       author = {{Albrecht}, Simon H. and {Marcussen}, Marcus L. and {Winn}, Joshua N. and {Dawson}, Rebekah I. and {Knudstrup}, Emil},
        title = "{A Preponderance of Perpendicular Planets}",
      journal = {\apjl},
         year = 2021,
        month = jul,
       volume = {916},
       number = {1},
          eid = {L1},
        pages = {L1},
          doi = {10.3847/2041-8213/ac0f03},
archivePrefix = {arXiv},
       eprint = {2105.09327},
 primaryClass = {astro-ph.EP},
       adsurl = {https://ui.adsabs.harvard.edu/abs/2021ApJ...916L...1A}
}

@ARTICLE{zak24,
       author = {{Zak}, J. and {Bocchieri}, A. and {Sedaghati}, E. and {Boffin}, H.~M.~J. and {Prudil}, Z. and {Skarka}, M. and {Changeat}, Q. and {Pascale}, E. and {Itrich}, D. and {Ivanov}, V.~D. and {Vitkova}, M. and {Kabath}, P. and {Roth}, M. and {Hatzes}, A.},
        title = "{Stellar obliquity measurements of six gas giants. Orbital misalignment of WASP-101b and WASP-131b}",
      journal = {\aap},
         year = 2024,
        month = jun,
       volume = {686},
          eid = {A147},
        pages = {A147},
          doi = {10.1051/0004-6361/202349084},
archivePrefix = {arXiv},
       eprint = {2403.15631},
 primaryClass = {astro-ph.EP},
       adsurl = {https://ui.adsabs.harvard.edu/abs/2024A&A...686A.147Z}
}

@ARTICLE{knud24,
       author = {{Knudstrup}, E. and {Albrecht}, S.~H. and {Winn}, J.~N. and {Gandolfi}, D. and {Zanazzi}, J.~J. and {Persson}, C.~M. and {Fridlund}, M. and {Marcussen}, M.~L. and {Chontos}, A. and {Keniger}, M.~A.~F. and {Eisner}, N.~L. and {Bieryla}, A. and {Isaacson}, H. and {Howard}, A.~W. and {Hirsch}, L.~A. and {Murgas}, F. and {Narita}, N. and {Palle}, E. and {Kawai}, Y. and {Baker}, D.},
        title = "{Obliquities of exoplanet host stars: Nineteen new and updated measurements, and trends in the sample of 205 measurements}",
      journal = {\aap},
         year = 2024,
        month = oct,
       volume = {690},
          eid = {A379},
        pages = {A379},
          doi = {10.1051/0004-6361/202450627},
archivePrefix = {arXiv},
       eprint = {2408.09793},
 primaryClass = {astro-ph.EP},
       adsurl = {https://ui.adsabs.harvard.edu/abs/2024A&A...690A.379K}
}

@ARTICLE{tin18,
       author = {{Tinetti}, Giovanna and {Drossart}, Pierre and {Eccleston}, Paul and {Hartogh}, Paul and {Heske}, Astrid and {Leconte}, J{\'e}r{\'e}my and {Micela}, Giusi and {Ollivier}, Marc and {Pilbratt}, G{\"o}ran and {Puig}, Ludovic and {Turrini}, Diego and {Vandenbussche}, Bart and {Wolkenberg}, Paulina and {Beaulieu}, Jean-Philippe and {Buchave}, Lars A. and {Ferus}, Martin and {Griffin}, Matt and {Guedel}, Manuel and {Justtanont}, Kay and {Lagage}, Pierre-Olivier and {Machado}, Pedro and {Malaguti}, Giuseppe and {Min}, Michiel and {N{\o}rgaard-Nielsen}, Hans Ulrik and {Rataj}, Mirek and {Ray}, Tom and {Ribas}, Ignasi and {Swain}, Mark and {Szabo}, Robert and {Werner}, Stephanie and {Barstow}, Joanna and {Burleigh}, Matt and {Cho}, James and {Coud{\'e} du Foresto}, Vincent and {Coustenis}, Athena and {Decin}, Leen and {Encrenaz}, Therese and {Galand}, Marina and {Gillon}, Michael and {Helled}, Ravit and {Morales}, Juan Carlos and {Garc{\'\i}a Mu{\~n}oz}, Antonio and {Moneti}, Andrea and {Pagano}, Isabella and {Pascale}, Enzo and {Piccioni}, Giuseppe and {Pinfield}, David and {Sarkar}, Subhajit and {Selsis}, Franck and {Tennyson}, Jonathan and {Triaud}, Amaury and {Venot}, Olivia and {Waldmann}, Ingo and {Waltham}, David and {Wright}, Gillian and {Amiaux}, Jerome and {Augu{\`e}res}, Jean-Louis and {Berth{\'e}}, Michel and {Bezawada}, Naidu and {Bishop}, Georgia and {Bowles}, Neil and {Coffey}, Deirdre and {Colom{\'e}}, Josep and {Crook}, Martin and {Crouzet}, Pierre-Elie and {Da Peppo}, Vania and {Sanz}, Isabel Escudero and {Focardi}, Mauro and {Frericks}, Martin and {Hunt}, Tom and {Kohley}, Ralf and {Middleton}, Kevin and {Morgante}, Gianluca and {Ottensamer}, Roland and {Pace}, Emanuele and {Pearson}, Chris and {Stamper}, Richard and {Symonds}, Kate and {Rengel}, Miriam and {Renotte}, Etienne and {Ade}, Peter and {Affer}, Laura and {Alard}, Christophe and {Allard}, Nicole and {Altieri}, Francesca and {Andr{\'e}}, Yves and {Arena}, Claudio and {Argyriou}, Ioannis and {Aylward}, Alan and {Baccani}, Cristian and {Bakos}, Gaspar and {Banaszkiewicz}, Marek and {Barlow}, Mike and {Batista}, Virginie and {Bellucci}, Giancarlo and {Benatti}, Serena and {Bernardi}, Pernelle and {B{\'e}zard}, Bruno and {Blecka}, Maria and {Bolmont}, Emeline and {Bonfond}, Bertrand and {Bonito}, Rosaria and {Bonomo}, Aldo S. and {Brucato}, John Robert and {Brun}, Allan Sacha and {Bryson}, Ian and {Bujwan}, Waldemar and {Casewell}, Sarah and {Charnay}, Bejamin and {Pestellini}, Cesare Cecchi and {Chen}, Guo and {Ciaravella}, Angela and {Claudi}, Riccardo and {Cl{\'e}dassou}, Rodolphe and {Damasso}, Mario and {Damiano}, Mario and {Danielski}, Camilla and {Deroo}, Pieter and {Di Giorgio}, Anna Maria and {Dominik}, Carsten and {Doublier}, Vanessa and {Doyle}, Simon and {Doyon}, Ren{\'e} and {Drummond}, Benjamin and {Duong}, Bastien and {Eales}, Stephen and {Edwards}, Billy and {Farina}, Maria and {Flaccomio}, Ettore and {Fletcher}, Leigh and {Forget}, Fran{\c{c}}ois and {Fossey}, Steve and {Fr{\"a}nz}, Markus and {Fujii}, Yuka and {Garc{\'\i}a-Piquer}, {\'A}lvaro and {Gear}, Walter and {Geoffray}, Herv{\'e} and {G{\'e}rard}, Jean Claude and {Gesa}, Lluis and {Gomez}, H. and {Graczyk}, Rafa{\l} and {Griffith}, Caitlin and {Grodent}, Denis and {Guarcello}, Mario Giuseppe and {Gustin}, Jacques and {Hamano}, Keiko and {Hargrave}, Peter and {Hello}, Yann and {Heng}, Kevin and {Herrero}, Enrique and {Hornstrup}, Allan and {Hubert}, Benoit and {Ida}, Shigeru and {Ikoma}, Masahiro and {Iro}, Nicolas and {Irwin}, Patrick and {Jarchow}, Christopher and {Jaubert}, Jean and {Jones}, Hugh and {Julien}, Queyrel and {Kameda}, Shingo and {Kerschbaum}, Franz and {Kervella}, Pierre and {Koskinen}, Tommi and {Krijger}, Matthijs and {Krupp}, Norbert and {Lafarga}, Marina and {Landini}, Federico and {Lellouch}, Emanuel and {Leto}, Giuseppe and {Luntzer}, A. and {Rank-L{\"u}ftinger}, Theresa and {Maggio}, Antonio and {Maldonado}, Jesus and {Maillard}, Jean-Pierre and {Mall}, Urs and {Marquette}, Jean-Baptiste and {Mathis}, Stephane and {Maxted}, Pierre and {Matsuo}, Taro and {Medvedev}, Alexander and {Miguel}, Yamila and {Minier}, Vincent and {Morello}, Giuseppe and {Mura}, Alessandro and {Narita}, Norio and {Nascimbeni}, Valerio and {Nguyen Tong}, N. and {Noce}, Vladimiro and {Oliva}, Fabrizio and {Palle}, Enric and {Palmer}, Paul and {Pancrazzi}, Maurizio and {Papageorgiou}, Andreas and {Parmentier}, Vivien and {Perger}, Manuel and {Petralia}, Antonino and {Pezzuto}, Stefano and {Pierrehumbert}, Ray and {Pillitteri}, Ignazio},
        title = "{A chemical survey of exoplanets with ARIEL}",
      journal = {Experimental Astronomy},
         year = 2018,
        month = nov,
       volume = {46},
       number = {1},
        pages = {135-209},
          doi = {10.1007/s10686-018-9598-x},
       adsurl = {https://ui.adsabs.harvard.edu/abs/2018ExA....46..135T}
}

@ARTICLE{kirk24,
       author = {{Kirk}, James and {Ahrer}, Eva-Maria and {Penzlin}, Anna B.~T. and {Owen}, James E. and {Booth}, Richard A. and {Alderson}, Lili and {Christie}, Duncan A. and {Claringbold}, Alastair B. and {Esparza-Borges}, Emma and {Fisher}, Chloe E. and {L{\'o}pez-Morales}, Mercedes and {Mayne}, N.~J. and {McCormack}, Mason and {Meech}, Annabella and {Panwar}, Vatsal and {Powell}, Diana and {Sergeev}, Denis E. and {Taylor}, Jake and {Tsai}, Shang-Min and {Valentine}, Daniel and {Wakeford}, Hannah R. and {Wheatley}, Peter J. and {Zamyatina}, Maria},
        title = "{BOWIE-ALIGN: A JWST comparative survey of aligned versus misaligned hot Jupiters to test the dependence of atmospheric composition on migration history}",
      journal = {RAS Techniques and Instruments},
         year = 2024,
        month = jan,
       volume = {3},
       number = {1},
        pages = {691-704},
          doi = {10.1093/rasti/rzae043},
archivePrefix = {arXiv},
       eprint = {2407.03198},
 primaryClass = {astro-ph.EP},
       adsurl = {https://ui.adsabs.harvard.edu/abs/2024RASTI...3..691K}
}

@ARTICLE{zak25b,
       author = {{Zak}, J. and {Boffin}, H.~M.~J. and {Bocchieri}, A. and {Sedaghati}, E. and {Balkoova}, Z. and {Kabath}, P.},
        title = "{Ten Aligned Orbits: Planet Migration in the Era of JWST and Ariel}",
      journal = {arXiv e-prints},
         year = 2025,
        month = may,
          eid = {arXiv:2505.20516},
        pages = {arXiv:2505.20516},
          doi = {10.48550/arXiv.2505.20516},
archivePrefix = {arXiv},
       eprint = {2505.20516},
 primaryClass = {astro-ph.EP},
       adsurl = {https://ui.adsabs.harvard.edu/abs/2025arXiv250520516Z}
}

@ARTICLE{edw22,
       author = {{Edwards}, Billy and {Tinetti}, Giovanna},
        title = "{The Ariel Target List: The Impact of TESS and the Potential for Characterizing Multiple Planets within a System}",
      journal = {\aj},
         year = 2022,
        month = jul,
       volume = {164},
       number = {1},
          eid = {15},
        pages = {15},
          doi = {10.3847/1538-3881/ac6bf9},
archivePrefix = {arXiv},
       eprint = {2205.05073},
 primaryClass = {astro-ph.EP},
       adsurl = {https://ui.adsabs.harvard.edu/abs/2022AJ....164...15E}
}

@ARTICLE{turr21,
       author = {{Turrini}, D. and {Schisano}, E. and {Fonte}, S. and {Molinari}, S. and {Politi}, R. and {Fedele}, D. and {Pani{\'c}}, O. and {Kama}, M. and {Changeat}, Q. and {Tinetti}, G.},
        title = "{Tracing the Formation History of Giant Planets in Protoplanetary Disks with Carbon, Oxygen, Nitrogen, and Sulfur}",
      journal = {\apj},
         year = 2021,
        month = mar,
       volume = {909},
       number = {1},
          eid = {40},
        pages = {40},
          doi = {10.3847/1538-4357/abd6e5},
archivePrefix = {arXiv},
       eprint = {2012.14315},
 primaryClass = {astro-ph.EP},
       adsurl = {https://ui.adsabs.harvard.edu/abs/2021ApJ...909...40T}
}

@ARTICLE{cow25,
       author = {{D'Aoust}, Lina and {Coull-Neveu}, Ben and {Lee}, Eve J. and {Cowan}, Nicolas B.},
        title = "{Testing the Origin of Hot Jupiters with Ariel}",
      journal = {arXiv e-prints},
         year = 2025,
        month = jul,
          eid = {arXiv:2507.13446},
        pages = {arXiv:2507.13446},
          doi = {10.48550/arXiv.2507.13446},
archivePrefix = {arXiv},
       eprint = {2507.13446},
 primaryClass = {astro-ph.EP},
       adsurl = {https://ui.adsabs.harvard.edu/abs/2025arXiv250713446D}
}

@INPROCEEDINGS{TNGarchive,
       author = {{Pasian}, Fabio},
        title = "{Archiving TNG Data}",
    booktitle = {Astronomical Data Analysis Software and Systems V},
         year = 1996,
       editor = {{Jacoby}, George H. and {Barnes}, Jeannette},
       series = {Astronomical Society of the Pacific Conference Series},
       volume = {101},
        month = jan,
        pages = {479},
       adsurl = {https://ui.adsabs.harvard.edu/abs/1996ASPC..101..479P}
}

@ARTICLE{YABI,
    author = {Hunter, Adam and Macgregor, Andrew and Szabo, Tamas and Wellington, Crispin and Bellgard, Matthew},
    year = {2012},
    month = {02},
    pages = {1},
    title = {Yabi: An online research environment for grid, high performance and cloud computing},
    volume = {7},
    journal = {Source code for biology and medicine},
    doi = {10.1186/1751-0473-7-1}
}

@ARTICLE{Pepe2002,
       author = {{Pepe}, F. and {Mayor}, M. and {Galland}, F. and {Naef}, D. and {Queloz}, D. and {Santos}, N.~C. and {Udry}, S. and {Burnet}, M.},
        title = "{The CORALIE survey for southern extra-solar planets VII. Two short-period Saturnian companions to <ASTROBJ>HD 108147</ASTROBJ> and <ASTROBJ>HD 168746</ASTROBJ>}",
      journal = {\aap},
         year = 2002,
        month = jun,
       volume = {388},
        pages = {632-638},
          doi = {10.1051/0004-6361:20020433},
archivePrefix = {arXiv},
       eprint = {astro-ph/0202457},
 primaryClass = {astro-ph},
       adsurl = {https://ui.adsabs.harvard.edu/abs/2002A&A...388..632P}
}

@ARTICLE{Ricker2015,
       author = {{Ricker}, George R. and {Winn}, Joshua N. and {Vanderspek}, Roland and {Latham}, David W. and {Bakos}, G{\'a}sp{\'a}r {\'A}. and {Bean}, Jacob L. and {Berta-Thompson}, Zachory K. and {Brown}, Timothy M. and {Buchhave}, Lars and {Butler}, Nathaniel R. and {Butler}, R. Paul and {Chaplin}, William J. and {Charbonneau}, David and {Christensen-Dalsgaard}, J{\o}rgen and {Clampin}, Mark and {Deming}, Drake and {Doty}, John and {De Lee}, Nathan and {Dressing}, Courtney and {Dunham}, Edward W. and {Endl}, Michael and {Fressin}, Francois and {Ge}, Jian and {Henning}, Thomas and {Holman}, Matthew J. and {Howard}, Andrew W. and {Ida}, Shigeru and {Jenkins}, Jon M. and {Jernigan}, Garrett and {Johnson}, John Asher and {Kaltenegger}, Lisa and {Kawai}, Nobuyuki and {Kjeldsen}, Hans and {Laughlin}, Gregory and {Levine}, Alan M. and {Lin}, Douglas and {Lissauer}, Jack J. and {MacQueen}, Phillip and {Marcy}, Geoffrey and {McCullough}, Peter R. and {Morton}, Timothy D. and {Narita}, Norio and {Paegert}, Martin and {Palle}, Enric and {Pepe}, Francesco and {Pepper}, Joshua and {Quirrenbach}, Andreas and {Rinehart}, Stephen A. and {Sasselov}, Dimitar and {Sato}, Bun'ei and {Seager}, Sara and {Sozzetti}, Alessandro and {Stassun}, Keivan G. and {Sullivan}, Peter and {Szentgyorgyi}, Andrew and {Torres}, Guillermo and {Udry}, Stephane and {Villasenor}, Joel},
        title = "{Transiting Exoplanet Survey Satellite (TESS)}",
      journal = {Journal of Astronomical Telescopes, Instruments, and Systems},
         year = 2015,
        month = jan,
       volume = {1},
          eid = {014003},
        pages = {014003},
          doi = {10.1117/1.JATIS.1.1.014003},
       adsurl = {https://ui.adsabs.harvard.edu/abs/2015JATIS...1a4003R}
}

@MISC{Rainer2013,
  author    = {Rainer, M.},
  title     = {Internal report GAPS-SCI-REP-007},
  year      = {2013},
  note      = {Internal report},
  howpublished = {GAPS Consortium},
}

@INPROCEEDINGS{BCuaresma2014,
       author = {{Blanco-Cuaresma}, S. and {Soubiran}, C. and {Jofr{\'e}}, P. and {Heiter}, U.},
        title = "{iSpec: An integrated software framework for the analysis of stellar spectra}",
    booktitle = {Astronomical Society of India Conference Series},
         year = 2014,
       series = {Astronomical Society of India Conference Series},
       volume = {11},
        month = jan,
        pages = {85-91},
archivePrefix = {arXiv},
       eprint = {1312.4545},
 primaryClass = {astro-ph.SR},
       adsurl = {https://ui.adsabs.harvard.edu/abs/2014ASInC..11...85B}
}

@ARTICLE{BCuaresma2019,
       author = {{Blanco-Cuaresma}, Sergi},
        title = "{Modern stellar spectroscopy caveats}",
      journal = {\mnras},
         year = 2019,
        month = jun,
       volume = {486},
       number = {2},
        pages = {2075-2101},
          doi = {10.1093/mnras/stz549},
archivePrefix = {arXiv},
       eprint = {1902.09558},
 primaryClass = {astro-ph.SR},
       adsurl = {https://ui.adsabs.harvard.edu/abs/2019MNRAS.486.2075B}
}

@INPROCEEDINGS{Jenkins2016,
       author = {{Jenkins}, Jon M. and {Twicken}, Joseph D. and {McCauliff}, Sean and {Campbell}, Jennifer and {Sanderfer}, Dwight and {Lung}, David and {Mansouri-Samani}, Masoud and {Girouard}, Forrest and {Tenenbaum}, Peter and {Klaus}, Todd and {Smith}, Jeffrey C. and {Caldwell}, Douglas A. and {Chacon}, A.~D. and {Henze}, Christopher and {Heiges}, Cory and {Latham}, David W. and {Morgan}, Edward and {Swade}, Daryl and {Rinehart}, Stephen and {Vanderspek}, Roland},
        title = "{The TESS science processing operations center}",
    booktitle = {Software and Cyberinfrastructure for Astronomy IV},
         year = 2016,
       editor = {{Chiozzi}, Gianluca and {Guzman}, Juan C.},
       series = {Society of Photo-Optical Instrumentation Engineers (SPIE) Conference Series},
       volume = {9913},
        month = aug,
          eid = {99133E},
        pages = {99133E},
          doi = {10.1117/12.2233418},
       adsurl = {https://ui.adsabs.harvard.edu/abs/2016SPIE.9913E..3EJ}
}

@software{SPECTRUM,
       author = {{Gray}, Richard O.},
        title = "{SPECTRUM: A stellar spectral synthesis program}",
 howpublished = {Astrophysics Source Code Library, record ascl:9910.002},
         year = 1999,
        month = oct,
          eid = {ascl:9910.002},
       adsurl = {https://ui.adsabs.harvard.edu/abs/1999ascl.soft10002G}
}

@INPROCEEDINGS{IRAF,
       author = {{Tody}, Doug},
        title = "{The IRAF Data Reduction and Analysis System}",
    booktitle = {Instrumentation in astronomy VI},
         year = 1986,
       editor = {{Crawford}, David L.},
       series = {Society of Photo-Optical Instrumentation Engineers (SPIE) Conference Series},
       volume = {627},
        month = jan,
        pages = {733},
          doi = {10.1117/12.968154},
       adsurl = {https://ui.adsabs.harvard.edu/abs/1986SPIE..627..733T}
}

@ARTICLE{Knudstrup2022,
       author = {{Knudstrup}, E. and {Albrecht}, S.~H.},
        title = "{Orbital alignment of HD 332231 b. The warm Saturn HD 332231 b/TOI-1456 b travels on a well-aligned, circular orbit around a bright F8 dwarf}",
      journal = {\aap},
         year = 2022,
        month = apr,
       volume = {660},
          eid = {A99},
        pages = {A99},
          doi = {10.1051/0004-6361/202142726},
archivePrefix = {arXiv},
       eprint = {2111.14968},
 primaryClass = {astro-ph.EP},
       adsurl = {https://ui.adsabs.harvard.edu/abs/2022A&A...660A..99K}
}

@ARTICLE{Dawson2018,
       author = {{Dawson}, Rebekah I. and {Johnson}, John Asher},
        title = "{Origins of Hot Jupiters}",
      journal = {\araa},
         year = 2018,
        month = sep,
       volume = {56},
        pages = {175-221},
          doi = {10.1146/annurev-astro-081817-051853},
archivePrefix = {arXiv},
       eprint = {1801.06117},
 primaryClass = {astro-ph.EP},
       adsurl = {https://ui.adsabs.harvard.edu/abs/2018ARA&A..56..175D}
}

@ARTICLE{Batygin2016,
       author = {{Batygin}, Konstantin and {Bodenheimer}, Peter H. and {Laughlin}, Gregory P.},
        title = "{In Situ Formation and Dynamical Evolution of Hot Jupiter Systems}",
      journal = {\apj},
         year = 2016,
        month = oct,
       volume = {829},
       number = {2},
          eid = {114},
        pages = {114},
          doi = {10.3847/0004-637X/829/2/114},
archivePrefix = {arXiv},
       eprint = {1511.09157},
 primaryClass = {astro-ph.EP},
       adsurl = {https://ui.adsabs.harvard.edu/abs/2016ApJ...829..114B}
}

@ARTICLE{Lin1986,
       author = {{Lin}, D.~N.~C. and {Papaloizou}, John},
        title = "{On the Tidal Interaction between Protoplanets and the Protoplanetary Disk. III. Orbital Migration of Protoplanets}",
      journal = {\apj},
         year = 1986,
        month = oct,
       volume = {309},
        pages = {846},
          doi = {10.1086/164653},
       adsurl = {https://ui.adsabs.harvard.edu/abs/1986ApJ...309..846L}
}

@ARTICLE{Murray1998,
       author = {{Murray}, N. and {Hansen}, B. and {Holman}, M. and {Tremaine}, S.},
        title = "{Migrating Planets}",
      journal = {Science},
         year = 1998,
        month = jan,
       volume = {279},
        pages = {69},
          doi = {10.1126/science.279.5347.69},
archivePrefix = {arXiv},
       eprint = {astro-ph/9801138},
 primaryClass = {astro-ph},
       adsurl = {https://ui.adsabs.harvard.edu/abs/1998Sci...279...69M}
}

@ARTICLE{Fabrycky2007,
       author = {{Fabrycky}, Daniel and {Tremaine}, Scott},
        title = "{Shrinking Binary and Planetary Orbits by Kozai Cycles with Tidal Friction}",
      journal = {\apj},
         year = 2007,
        month = nov,
       volume = {669},
       number = {2},
        pages = {1298-1315},
          doi = {10.1086/521702},
archivePrefix = {arXiv},
       eprint = {0705.4285},
 primaryClass = {astro-ph},
       adsurl = {https://ui.adsabs.harvard.edu/abs/2007ApJ...669.1298F}
}

@ARTICLE{Giacalone2017,
       author = {{Giacalone}, Steven and {Matsakos}, Titos and {K{\"o}nigl}, Arieh},
        title = "{A Test of the High-eccentricity Migration Scenario for Close-in Planets}",
      journal = {\aj},
         year = 2017,
        month = nov,
       volume = {154},
       number = {5},
          eid = {192},
        pages = {192},
          doi = {10.3847/1538-3881/aa8c04},
archivePrefix = {arXiv},
       eprint = {1708.07543},
 primaryClass = {astro-ph.EP},
       adsurl = {https://ui.adsabs.harvard.edu/abs/2017AJ....154..192G}
}

@ARTICLE{Brown2012,
       author = {{Brown}, D.~J.~A. and {Collier Cameron}, A. and {D{\'\i}az}, R.~F. and {Doyle}, A.~P. and {Gillon}, M. and {Lendl}, M. and {Smalley}, B. and {Triaud}, A.~H.~M.~J. and {Anderson}, D.~R. and {Enoch}, B. and {Hellier}, C. and {Maxted}, P.~F.~L. and {Miller}, G.~R.~M. and {Pollacco}, D. and {Queloz}, D. and {Boisse}, I. and {H{\'e}brard}, G.},
        title = "{Analysis of Spin-Orbit Alignment in the WASP-32, WASP-38, and HAT-P-27/WASP-40 Systems}",
      journal = {\apj},
         year = 2012,
        month = dec,
       volume = {760},
       number = {2},
          eid = {139},
        pages = {139},
          doi = {10.1088/0004-637X/760/2/139},
archivePrefix = {arXiv},
       eprint = {1303.5649},
 primaryClass = {astro-ph.EP},
       adsurl = {https://ui.adsabs.harvard.edu/abs/2012ApJ...760..139B}
}

@ARTICLE{Bourrier2023,
       author = {{Bourrier}, V. and {Attia}, M. and {Mallonn}, M. and {Marret}, A. and {Lendl}, M. and {Konig}, P. -C. and {Krenn}, A. and {Cretignier}, M. and {Allart}, R. and {Henry}, G. and {Bryant}, E. and {Leleu}, A. and {Nielsen}, L. and {Hebrard}, G. and {Hara}, N. and {Ehrenreich}, D. and {Seidel}, J. and {dos Santos}, L. and {Lovis}, C. and {Bayliss}, D. and {Cegla}, H.~M. and {Dumusque}, X. and {Boisse}, I. and {Boucher}, A. and {Bouchy}, F. and {Pepe}, F. and {Lavie}, B. and {Rey Cerda}, J. and {S{\'e}gransan}, D. and {Udry}, S. and {Vrignaud}, T.},
        title = "{DREAM: I. Orbital architecture orrery}",
      journal = {\aap},
         year = 2023,
        month = jan,
       volume = {669},
          eid = {A63},
        pages = {A63},
          doi = {10.1051/0004-6361/202245004},
archivePrefix = {arXiv},
       eprint = {2301.07727},
 primaryClass = {astro-ph.EP},
       adsurl = {https://ui.adsabs.harvard.edu/abs/2023A&A...669A..63B}
}

@ARTICLE{Bonomo2017,
       author = {{Bonomo}, A.~S. and {Desidera}, S. and {Benatti}, S. and {Borsa}, F. and {Crespi}, S. and {Damasso}, M. and {Lanza}, A.~F. and {Sozzetti}, A. and {Lodato}, G. and {Marzari}, F. and {Boccato}, C. and {Claudi}, R.~U. and {Cosentino}, R. and {Covino}, E. and {Gratton}, R. and {Maggio}, A. and {Micela}, G. and {Molinari}, E. and {Pagano}, I. and {Piotto}, G. and {Poretti}, E. and {Smareglia}, R. and {Affer}, L. and {Biazzo}, K. and {Bignamini}, A. and {Esposito}, M. and {Giacobbe}, P. and {H{\'e}brard}, G. and {Malavolta}, L. and {Maldonado}, J. and {Mancini}, L. and {Martinez Fiorenzano}, A. and {Masiero}, S. and {Nascimbeni}, V. and {Pedani}, M. and {Rainer}, M. and {Scandariato}, G.},
        title = "{The GAPS Programme with HARPS-N at TNG . XIV. Investigating giant planet migration history via improved eccentricity and mass determination for 231 transiting planets}",
      journal = {\aap},
         year = 2017,
        month = jun,
       volume = {602},
          eid = {A107},
        pages = {A107},
          doi = {10.1051/0004-6361/201629882},
archivePrefix = {arXiv},
       eprint = {1704.00373},
 primaryClass = {astro-ph.EP},
       adsurl = {https://ui.adsabs.harvard.edu/abs/2017A&A...602A.107B}
}

@ARTICLE{Hartman2015,
       author = {{Hartman}, J.~D. and {Bakos}, G. {\'A}. and {Buchhave}, L.~A. and {Torres}, G. and {Latham}, D.~W. and {Kov{\'a}cs}, G. and {Bhatti}, W. and {Csubry}, Z. and {de Val-Borro}, M. and {Penev}, K. and {Huang}, C.~X. and {B{\'e}ky}, B. and {Bieryla}, A. and {Quinn}, S.~N. and {Howard}, A.~W. and {Marcy}, G.~W. and {Johnson}, J.~A. and {Isaacson}, H. and {Fischer}, D.~A. and {Noyes}, R.~W. and {Falco}, E. and {Esquerdo}, G.~A. and {Knox}, R.~P. and {Hinz}, P. and {L{\'a}z{\'a}r}, J. and {Papp}, I. and {S{\'a}ri}, P.},
        title = "{HAT-P-57b: A Short-period Giant Planet Transiting a Bright Rapidly Rotating A8V Star Confirmed Via Doppler Tomography}",
      journal = {\aj},
         year = 2015,
        month = dec,
       volume = {150},
       number = {6},
          eid = {197},
        pages = {197},
          doi = {10.1088/0004-6256/150/6/197},
archivePrefix = {arXiv},
       eprint = {1510.08839},
 primaryClass = {astro-ph.EP},
       adsurl = {https://ui.adsabs.harvard.edu/abs/2015AJ....150..197H}
}

@ARTICLE{Wong2014,
       author = {{Wong}, Ian and {Knutson}, Heather A. and {Cowan}, Nicolas B. and {Lewis}, Nikole K. and {Agol}, Eric and {Burrows}, Adam and {Deming}, Drake and {Fortney}, Jonathan J. and {Fulton}, Benjamin J. and {Langton}, Jonathan and {Laughlin}, Gregory and {Showman}, Adam P.},
        title = "{Constraints on the Atmospheric Circulation and Variability of the Eccentric Hot Jupiter XO-3b}",
      journal = {\apj},
         year = 2014,
        month = oct,
       volume = {794},
       number = {2},
          eid = {134},
        pages = {134},
          doi = {10.1088/0004-637X/794/2/134},
archivePrefix = {arXiv},
       eprint = {1407.1313},
 primaryClass = {astro-ph.EP},
       adsurl = {https://ui.adsabs.harvard.edu/abs/2014ApJ...794..134W}
}

@ARTICLE{Winn2008,
       author = {{Winn}, Joshua N. and {Holman}, Matthew J. and {Torres}, Guillermo and {McCullough}, Peter and {Johns-Krull}, Christopher and {Latham}, David W. and {Shporer}, Avi and {Mazeh}, Tsevi and {Garcia-Melendo}, Enrique and {Foote}, Cindy and {Esquerdo}, Gil and {Everett}, Mark},
        title = "{The Transit Light Curve Project. IX. Evidence for a Smaller Radius of the Exoplanet XO-3b}",
      journal = {\apj},
         year = 2008,
        month = aug,
       volume = {683},
       number = {2},
        pages = {1076-1084},
          doi = {10.1086/589737},
archivePrefix = {arXiv},
       eprint = {0804.4475},
 primaryClass = {astro-ph},
       adsurl = {https://ui.adsabs.harvard.edu/abs/2008ApJ...683.1076W}
}

@ARTICLE{Winn2009,
       author = {{Winn}, Joshua N. and {Johnson}, John Asher and {Fabrycky}, Daniel and {Howard}, Andrew W. and {Marcy}, Geoffrey W. and {Narita}, Norio and {Crossfield}, Ian J. and {Suto}, Yasushi and {Turner}, Edwin L. and {Esquerdo}, Gil and {Holman}, Matthew J.},
        title = "{On the Spin-Orbit Misalignment of the XO-3 Exoplanetary System}",
      journal = {\apj},
         year = 2009,
        month = jul,
       volume = {700},
       number = {1},
        pages = {302-308},
          doi = {10.1088/0004-637X/700/1/302},
archivePrefix = {arXiv},
       eprint = {0902.3461},
 primaryClass = {astro-ph.EP},
       adsurl = {https://ui.adsabs.harvard.edu/abs/2009ApJ...700..302W}
}

@ARTICLE{Winn2010,
       author = {{Winn}, Joshua N. and {Fabrycky}, Daniel and {Albrecht}, Simon and {Johnson}, John Asher},
        title = "{Hot Stars with Hot Jupiters Have High Obliquities}",
      journal = {\apjl},
         year = 2010,
        month = aug,
       volume = {718},
       number = {2},
        pages = {L145-L149},
          doi = {10.1088/2041-8205/718/2/L145},
archivePrefix = {arXiv},
       eprint = {1006.4161},
 primaryClass = {astro-ph.EP},
       adsurl = {https://ui.adsabs.harvard.edu/abs/2010ApJ...718L.145W}
}

@ARTICLE{Southworth2010,
       author = {{Southworth}, John},
        title = "{Homogeneous studies of transiting extrasolar planets - III. Additional planets and stellar models}",
      journal = {\mnras},
         year = 2010,
        month = nov,
       volume = {408},
       number = {3},
        pages = {1689-1713},
          doi = {10.1111/j.1365-2966.2010.17231.x},
archivePrefix = {arXiv},
       eprint = {1006.4443},
 primaryClass = {astro-ph.EP},
       adsurl = {https://ui.adsabs.harvard.edu/abs/2010MNRAS.408.1689S}
}

@ARTICLE{Rossiter1924,
       author = {{Rossiter}, R.~A.},
        title = "{On the detection of an effect of rotation during eclipse in the velocity of the brigher component of beta Lyrae, and on the constancy of velocity of this system.}",
      journal = {\apj},
         year = 1924,
        month = jul,
       volume = {60},
        pages = {15-21},
          doi = {10.1086/142825},
       adsurl = {https://ui.adsabs.harvard.edu/abs/1924ApJ....60...15R}
}

@ARTICLE{McLaughlin1924,
       author = {{McLaughlin}, D.~B.},
        title = "{Some results of a spectrographic study of the Algol system.}",
      journal = {\apj},
         year = 1924,
        month = jul,
       volume = {60},
        pages = {22-31},
          doi = {10.1086/142826},
       adsurl = {https://ui.adsabs.harvard.edu/abs/1924ApJ....60...22M}
}

@ARTICLE{CollierCameron2010A,
       author = {{Collier Cameron}, A. and {Bruce}, V.~A. and {Miller}, G.~R.~M. and {Triaud}, A.~H.~M.~J. and {Queloz}, D.},
        title = "{Line-profile tomography of exoplanet transits - I. The Doppler shadow of HD 189733b}",
      journal = {\mnras},
         year = 2010,
        month = mar,
       volume = {403},
       number = {1},
        pages = {151-158},
          doi = {10.1111/j.1365-2966.2009.16131.x},
archivePrefix = {arXiv},
       eprint = {0911.5361},
 primaryClass = {astro-ph.SR},
       adsurl = {https://ui.adsabs.harvard.edu/abs/2010MNRAS.403..151C}
}

@ARTICLE{CollierCameron2010B,
       author = {{Collier Cameron}, A. and {Guenther}, E. and {Smalley}, B. and {McDonald}, I. and {Hebb}, L. and {Andersen}, J. and {Augusteijn}, Th. and {Barros}, S.~C.~C. and {Brown}, D.~J.~A. and {Cochran}, W.~D. and {Endl}, M. and {Fossey}, S.~J. and {Hartmann}, M. and {Maxted}, P.~F.~L. and {Pollacco}, D. and {Skillen}, I. and {Telting}, J. and {Waldmann}, I.~P. and {West}, R.~G.},
        title = "{Line-profile tomography of exoplanet transits - II. A gas-giant planet transiting a rapidly rotating A5 star}",
      journal = {\mnras},
         year = 2010,
        month = sep,
       volume = {407},
       number = {1},
        pages = {507-514},
          doi = {10.1111/j.1365-2966.2010.16922.x},
archivePrefix = {arXiv},
       eprint = {1004.4551},
 primaryClass = {astro-ph.EP},
       adsurl = {https://ui.adsabs.harvard.edu/abs/2010MNRAS.407..507C}
}

@ARTICLE{Zhou2016,
       author = {{Zhou}, George and {Rodriguez}, Joseph E. and {Collins}, Karen A. and {Beatty}, Thomas and {Oberst}, Thomas and {Heintz}, Tyler M. and {Stassun}, Keivan G. and {Latham}, David W. and {Kuhn}, Rudolf B. and {Bieryla}, Allyson and {Lund}, Michael B. and {Labadie-Bartz}, Jonathan and {Siverd}, Robert J. and {Stevens}, Daniel J. and {Gaudi}, B. Scott and {Pepper}, Joshua and {Buchhave}, Lars A. and {Eastman}, Jason and {Col{\'o}n}, Knicole and {Cargile}, Phillip and {James}, David and {Gregorio}, Joao and {Reed}, Phillip A. and {Jensen}, Eric L.~N. and {Cohen}, David H. and {McLeod}, Kim K. and {Tan}, T.~G. and {Zambelli}, Roberto and {Bayliss}, Daniel and {Bento}, Joao and {Esquerdo}, Gilbert A. and {Berlind}, Perry and {Calkins}, Michael L. and {Blancato}, Kirsten and {Manner}, Mark and {Samulski}, Camile and {Stockdale}, Christopher and {Nelson}, Peter and {Stephens}, Denise and {Curtis}, Ivan and {Kielkopf}, John and {Fulton}, Benjamin J. and {DePoy}, D.~L. and {Marshall}, Jennifer L. and {Pogge}, Richard and {Gould}, Andy and {Trueblood}, Mark and {Trueblood}, Pat},
        title = "{KELT-17b: A Hot-Jupiter Transiting an A-star in a Misaligned Orbit Detected with Doppler Tomography}",
      journal = {\aj},
         year = 2016,
        month = nov,
       volume = {152},
       number = {5},
          eid = {136},
        pages = {136},
          doi = {10.3847/0004-6256/152/5/136},
archivePrefix = {arXiv},
       eprint = {1607.03512},
 primaryClass = {astro-ph.EP},
       adsurl = {https://ui.adsabs.harvard.edu/abs/2016AJ....152..136Z}
}

@ARTICLE{Albrecht2013,
       author = {{Albrecht}, Simon and {Winn}, Joshua N. and {Marcy}, Geoffrey W. and {Howard}, Andrew W. and {Isaacson}, Howard and {Johnson}, John A.},
        title = "{Low Stellar Obliquities in Compact Multiplanet Systems}",
      journal = {\apj},
         year = 2013,
        month = jul,
       volume = {771},
       number = {1},
          eid = {11},
        pages = {11},
          doi = {10.1088/0004-637X/771/1/11},
archivePrefix = {arXiv},
       eprint = {1302.4443},
 primaryClass = {astro-ph.SR},
       adsurl = {https://ui.adsabs.harvard.edu/abs/2013ApJ...771...11A}
}

@article{Albrecht2022,
    doi = {10.1088/1538-3873/ac6c09},
    url = {https://dx.doi.org/10.1088/1538-3873/ac6c09},
    year = {2022},
    month = {aug},
    publisher = {The Astronomical Society of the Pacific},
    volume = {134},
    number = {1038},
    pages = {082001},
    author = {Albrecht, Simon H. and Dawson, Rebekah I. and Winn, Joshua N.},
    title = {Stellar Obliquities in Exoplanetary Systems},
    journal = {Publications of the Astronomical Society of the Pacific}
}

@ARTICLE{Johnson2014,
       author = {{Johnson}, Marshall C. and {Cochran}, William D. and {Albrecht}, Simon and {Dodson-Robinson}, Sarah E. and {Winn}, Joshua N. and {Gullikson}, Kevin},
        title = "{A Misaligned Prograde Orbit for Kepler-13 Ab via Doppler Tomography}",
      journal = {\apj},
         year = 2014,
        month = jul,
       volume = {790},
       number = {1},
          eid = {30},
        pages = {30},
          doi = {10.1088/0004-637X/790/1/30},
archivePrefix = {arXiv},
       eprint = {1406.0512},
 primaryClass = {astro-ph.EP},
       adsurl = {https://ui.adsabs.harvard.edu/abs/2014ApJ...790...30J}
}

@ARTICLE{Bieryla2014,
       author = {{Bieryla}, A. and {Hartman}, J.~D. and {Bakos}, G. {\'A}. and {Bhatti}, W. and {Kov{\'a}cs}, G. and {Boisse}, I. and {Latham}, D.~W. and {Buchhave}, L.~A. and {Csubry}, Z. and {Penev}, K. and {de Val-Borro}, M. and {B{\'e}ky}, B. and {Falco}, E. and {Torres}, G. and {Noyes}, R.~W. and {Berlind}, P. and {Calkins}, M.~C. and {Esquerdo}, G.~A. and {L{\'a}z{\'a}r}, J. and {Papp}, I. and {S{\'a}ri}, P.},
        title = "{HAT-P-49b: A 1.7 M $_{J}$ Planet Transiting a Bright 1.5 M $_{{\ensuremath{\odot}}}$ F-star}",
      journal = {\aj},
         year = 2014,
        month = apr,
       volume = {147},
       number = {4},
          eid = {84},
        pages = {84},
          doi = {10.1088/0004-6256/147/4/84},
archivePrefix = {arXiv},
       eprint = {1401.5460},
 primaryClass = {astro-ph.EP},
       adsurl = {https://ui.adsabs.harvard.edu/abs/2014AJ....147...84B}
}

@ARTICLE{Grevesse1998,
       author = {{Grevesse}, N. and {Sauval}, A.~J.},
        title = "{Standard Solar Composition}",
      journal = {\ssr},
         year = 1998,
        month = may,
       volume = {85},
        pages = {161-174},
          doi = {10.1023/A:1005161325181},
       adsurl = {https://ui.adsabs.harvard.edu/abs/1998SSRv...85..161G}
}

@ARTICLE{Plez2008,
       author = {{Plez}, B.},
        title = "{MARCS model atmospheres}",
      journal = {Physica Scripta Volume T},
         year = 2008,
        month = dec,
       volume = {133},
          eid = {014003},
        pages = {014003},
          doi = {10.1088/0031-8949/2008/T133/014003},
archivePrefix = {arXiv},
       eprint = {0810.2375},
 primaryClass = {astro-ph},
       adsurl = {https://ui.adsabs.harvard.edu/abs/2008PhST..133a4003P}
}

@ARTICLE{Hebrard2008,
       author = {{H{\'e}brard}, G. and {Bouchy}, F. and {Pont}, F. and {Loeillet}, B. and {Rabus}, M. and {Bonfils}, X. and {Moutou}, C. and {Boisse}, I. and {Delfosse}, X. and {Desort}, M. and {Eggenberger}, A. and {Ehrenreich}, D. and {Forveille}, T. and {Lagrange}, A. -M. and {Lovis}, C. and {Mayor}, M. and {Pepe}, F. and {Perrier}, C. and {Queloz}, D. and {Santos}, N.~C. and {S{\'e}gransan}, D. and {Udry}, S. and {Vidal-Madjar}, A.},
        title = "{Misaligned spin-orbit in the XO-3 planetary system?}",
      journal = {\aap},
         year = 2008,
        month = sep,
       volume = {488},
       number = {2},
        pages = {763-770},
          doi = {10.1051/0004-6361:200810056},
archivePrefix = {arXiv},
       eprint = {0806.0719},
 primaryClass = {astro-ph},
       adsurl = {https://ui.adsabs.harvard.edu/abs/2008A&A...488..763H}
}

@ARTICLE{Johns-Krull2008,
       author = {{Johns-Krull}, Christopher M. and {McCullough}, Peter R. and {Burke}, Christopher J. and {Valenti}, Jeff A. and {Janes}, K.~A. and {Heasley}, J.~N. and {Prato}, L. and {Bissinger}, R. and {Fleenor}, M. and {Foote}, C.~N. and {Garcia-Melendo}, E. and {Gary}, B.~L. and {Howell}, P.~J. and {Mallia}, F. and {Masi}, G. and {Vanmunster}, T.},
        title = "{XO-3b: A Massive Planet in an Eccentric Orbit Transiting an F5 V Star}",
      journal = {\apj},
         year = 2008,
        month = apr,
       volume = {677},
       number = {1},
        pages = {657-670},
          doi = {10.1086/528950},
archivePrefix = {arXiv},
       eprint = {0712.4283},
 primaryClass = {astro-ph},
       adsurl = {https://ui.adsabs.harvard.edu/abs/2008ApJ...677..657J}
}

@ARTICLE{Stassun2017,
       author = {{Stassun}, Keivan G. and {Collins}, Karen A. and {Gaudi}, B. Scott},
        title = "{Accurate Empirical Radii and Masses of Planets and Their Host Stars with Gaia Parallaxes}",
      journal = {\aj},
         year = 2017,
        month = mar,
       volume = {153},
       number = {3},
          eid = {136},
        pages = {136},
          doi = {10.3847/1538-3881/aa5df3},
archivePrefix = {arXiv},
       eprint = {1609.04389},
 primaryClass = {astro-ph.EP},
       adsurl = {https://ui.adsabs.harvard.edu/abs/2017AJ....153..136S}
}

@ARTICLE{Hirano2011,
       author = {{Hirano}, Teruyuki and {Narita}, Norio and {Sato}, Bun'ei and {Winn}, Joshua N. and {Aoki}, Wako and {Tamura}, Motohide and {Taruya}, Atsushi and {Suto}, Yasushi},
        title = "{Further Observations of the Tilted Planet XO-3: A New Determination of Spin-Orbit Misalignment, and Limits on Differential Rotation}",
      journal = {\pasj},
         year = 2011,
        month = dec,
       volume = {63},
       number = {6},
        pages = {L57-L61},
          doi = {10.1093/pasj/63.6.L57},
archivePrefix = {arXiv},
       eprint = {1108.4493},
 primaryClass = {astro-ph.EP},
       adsurl = {https://ui.adsabs.harvard.edu/abs/2011PASJ...63L..57H}
}

@ARTICLE{Rusznak2025,
       author = {{Rusznak}, Jace and {Wang}, Xian-Yu and {Rice}, Malena and {Wang}, Songhu},
        title = "{From Misaligned Sub-Saturns to Aligned Brown Dwarfs: The Highest M$_{p}$/M$_{*}$ Systems Exhibit Low Obliquities, Even around Hot Stars}",
      journal = {\apjl},
         year = 2025,
        month = apr,
       volume = {983},
       number = {2},
          eid = {L42},
        pages = {L42},
          doi = {10.3847/2041-8213/adc129},
archivePrefix = {arXiv},
       eprint = {2412.04438},
 primaryClass = {astro-ph.EP},
       adsurl = {https://ui.adsabs.harvard.edu/abs/2025ApJ...983L..42R}
}

@ARTICLE{Doyle2014,
       author = {{Doyle}, Amanda P. and {Davies}, Guy R. and {Smalley}, Barry and {Chaplin}, William J. and {Elsworth}, Yvonne},
        title = "{Determining stellar macroturbulence using asteroseismic rotational velocities from Kepler}",
      journal = {\mnras},
         year = 2014,
        month = nov,
       volume = {444},
       number = {4},
        pages = {3592-3602},
          doi = {10.1093/mnras/stu1692},
archivePrefix = {arXiv},
       eprint = {1408.3988},
 primaryClass = {astro-ph.SR},
       adsurl = {https://ui.adsabs.harvard.edu/abs/2014MNRAS.444.3592D}
}

@article{Kochukhov2007,
    author = {Kochukhov, O. and Ryabchikova, T. and Weiss, W. W. and Landstreet, J. D. and Lyashko, D.},
    title = {Line profile variations in rapidly oscillating Ap stars: resolution of the enigma*},
    journal = {Monthly Notices of the Royal Astronomical Society},
    volume = {376},
    number = {2},
    pages = {651-672},
    year = {2007},
    month = {04},
    issn = {0035-8711},
    doi = {10.1111/j.1365-2966.2007.11449.x},
    url = {https://doi.org/10.1111/j.1365-2966.2007.11449.x},
    eprint = {https://academic.oup.com/mnras/article-pdf/376/2/651/18670752/mnras0376-0651.pdf},
}

@article{KovariZsoltBartus2007,
    author = {Kovari, Zsolt and Bartus, J. and Strassmeier, K. and Oláh, K. and Weber, M. and Rice, John and Washuettl, A.},
    year = {2007},
    month = {03},
    pages = {},
    title = {Doppler imaging of stellar surface structure},
    volume = {373},
    journal = {http://dx.doi.org/10.1051/0004-6361:20065982},
    doi = {10.1051/0004-6361:20065982}
}

@ARTICLE{Rice2022,
       author = {{Rice}, Malena and {Wang}, Songhu and {Laughlin}, Gregory},
        title = "{Origins of Hot Jupiters from the Stellar Obliquity Distribution}",
      journal = {\apjl},
         year = 2022,
        month = feb,
       volume = {926},
       number = {2},
          eid = {L17},
        pages = {L17},
          doi = {10.3847/2041-8213/ac502d},
archivePrefix = {arXiv},
       eprint = {2201.11768},
 primaryClass = {astro-ph.EP},
       adsurl = {https://ui.adsabs.harvard.edu/abs/2022ApJ...926L..17R}
}

@ARTICLE{Albrecht2012,
       author = {{Albrecht}, Simon and {Winn}, Joshua N. and {Johnson}, John A. and {Howard}, Andrew W. and {Marcy}, Geoffrey W. and {Butler}, R. Paul and {Arriagada}, Pamela and {Crane}, Jeffrey D. and {Shectman}, Stephen A. and {Thompson}, Ian B. and {Hirano}, Teruyuki and {Bakos}, Gaspar and {Hartman}, Joel D.},
        title = "{Obliquities of Hot Jupiter Host Stars: Evidence for Tidal Interactions and Primordial Misalignments}",
      journal = {\apj},
         year = 2012,
        month = sep,
       volume = {757},
       number = {1},
          eid = {18},
        pages = {18},
          doi = {10.1088/0004-637X/757/1/18},
archivePrefix = {arXiv},
       eprint = {1206.6105},
 primaryClass = {astro-ph.SR},
       adsurl = {https://ui.adsabs.harvard.edu/abs/2012ApJ...757...18A}
}

@ARTICLE{Kraft1967,
       author = {{Kraft}, Robert P.},
        title = "{Studies of Stellar Rotation. V. The Dependence of Rotation on Age among Solar-Type Stars}",
      journal = {\apj},
         year = 1967,
        month = nov,
       volume = {150},
        pages = {551},
          doi = {10.1086/149359},
       adsurl = {https://ui.adsabs.harvard.edu/abs/1967ApJ...150..551K}
}

@ARTICLE{Smith2013,
       author = {{Smith}, A.~M.~S. and {Anderson}, D.~R. and {Bouchy}, F. and {Collier Cameron}, A. and {Doyle}, A.~P. and {Fumel}, A. and {Gillon}, M. and {H{\'e}brard}, G. and {Hellier}, C. and {Jehin}, E. and {Lendl}, M. and {Maxted}, P.~F.~L. and {Moutou}, C. and {Pepe}, F. and {Pollacco}, D. and {Queloz}, D. and {Santerne}, A. and {Segransan}, D. and {Smalley}, B. and {Southworth}, J. and {Triaud}, A.~H.~M.~J. and {Udry}, S. and {West}, R.~G.},
        title = "{WASP-71b: a bloated hot Jupiter in a 2.9-day, prograde orbit around an evolved F8 star}",
      journal = {\aap},
         year = 2013,
        month = apr,
       volume = {552},
          eid = {A120},
        pages = {A120},
          doi = {10.1051/0004-6361/201220727},
archivePrefix = {arXiv},
       eprint = {1211.3045},
 primaryClass = {astro-ph.EP},
       adsurl = {https://ui.adsabs.harvard.edu/abs/2013A&A...552A.120S}
}

@ARTICLE{Brown2017,
       author = {{Brown}, D.~J.~A. and {Triaud}, A.~H.~M.~J. and {Doyle}, A.~P. and {Gillon}, M. and {Lendl}, M. and {Anderson}, D.~R. and {Collier Cameron}, A. and {H{\'e}brard}, G. and {Hellier}, C. and {Lovis}, C. and {Maxted}, P.~F.~L. and {Pepe}, F. and {Pollacco}, D. and {Queloz}, D. and {Smalley}, B.},
        title = "{Rossiter-McLaughlin models and their effect on estimates of stellar rotation, illustrated using six WASP systems}",
      journal = {\mnras},
         year = 2017,
        month = jan,
       volume = {464},
       number = {1},
        pages = {810-839},
          doi = {10.1093/mnras/stw2316},
archivePrefix = {arXiv},
       eprint = {1610.00600},
 primaryClass = {astro-ph.EP},
       adsurl = {https://ui.adsabs.harvard.edu/abs/2017MNRAS.464..810B}
}

@ARTICLE{Sedaghati2022,
       author = {{Sedaghati}, E. and {S{\'a}nchez-L{\'o}pez}, A. and {Czesla}, S. and {L{\'o}pez-Puertas}, M. and {Amado}, P.~J. and {Palle}, E. and {Molaverdikhani}, K. and {Caballero}, J.~A. and {Nortmann}, L. and {Quirrenbach}, A. and {Reiners}, A. and {Ribas}, I.},
        title = "{Moderately misaligned orbit of the warm sub-Saturn HD 332231 b}",
      journal = {\aap},
         year = 2022,
        month = mar,
       volume = {659},
          eid = {A44},
        pages = {A44},
          doi = {10.1051/0004-6361/202142471},
archivePrefix = {arXiv},
       eprint = {2110.10282},
 primaryClass = {astro-ph.EP},
       adsurl = {https://ui.adsabs.harvard.edu/abs/2022A&A...659A..44S}
}

@article{Bate2010,
    author = {Bate, M. R. and Lodato, G. and Pringle, J. E.},
    title = {Chaotic star formation and the alignment of stellar rotation with disc and planetary orbital axes},
    journal = {Monthly Notices of the Royal Astronomical Society},
    volume = {401},
    number = {3},
    pages = {1505-1513},
    year = {2010},
    month = {01},
    issn = {0035-8711},
    doi = {10.1111/j.1365-2966.2009.15773.x},
    url = {https://doi.org/10.1111/j.1365-2966.2009.15773.x},
    eprint = {https://academic.oup.com/mnras/article-pdf/401/3/1505/3806179/mnras0401-1505.pdf},
}

@article{Lai2011,
    author = {Lai, Dong and Foucart, Francois and Lin, Douglas N. C.},
    title = {Evolution of spin direction of accreting magnetic protostars and spin–orbit misalignment in exoplanetary systems},
    journal = {Monthly Notices of the Royal Astronomical Society},
    volume = {412},
    number = {4},
    pages = {2790-2798},
    year = {2011},
    month = {04},
    issn = {0035-8711},
    doi = {10.1111/j.1365-2966.2010.18127.x},
    url = {https://doi.org/10.1111/j.1365-2966.2010.18127.x},
    eprint = {https://academic.oup.com/mnras/article-pdf/412/4/2790/3375437/mnras0412-2790.pdf},
}

@ARTICLE{Uytterhoeven2008,
       author = {{Uytterhoeven}, K. and {Mathias}, P. and {Poretti}, E. and {Rainer}, M. and {Mart{\'\i}n-Ruiz}, S. and {Rodr{\'\i}guez}, E. and {Amado}, P.~J. and {Le Contel}, D. and {Jankov}, S. and {Niemczura}, E. and {Pollard}, K.~R. and {Brunsden}, E. and {Papar{\'o}}, M. and {Costa}, V. and {Valtier}, J. -C. and {Garrido}, R. and {Su{\'a}rez}, J.~C. and {Kilmartin}, P.~M. and {Chapellier}, E. and {Rodr{\'\i}guez-L{\'o}pez}, C. and {Marin}, A.~J. and {Aceituno}, F.~J. and {Casanova}, V. and {Rolland}, A. and {Olivares}, I.},
        title = "{The {\ensuremath{\gamma}} Doradus CoRoT target HD 49434. I. Results from the ground-based campaign}",
      journal = {\aap},
         year = 2008,
        month = oct,
       volume = {489},
       number = {3},
        pages = {1213-1224},
          doi = {10.1051/0004-6361:200809992},
archivePrefix = {arXiv},
       eprint = {0807.0904},
 primaryClass = {astro-ph},
       adsurl = {https://ui.adsabs.harvard.edu/abs/2008A&A...489.1213U}
}

@ARTICLE{Mathias2004,
       author = {{Mathias}, P. and {Le Contel}, J. -M. and {Chapellier}, E. and {Jankov}, S. and {Sareyan}, J. -P. and {Poretti}, E. and {Garrido}, R. and {Rodr{\'\i}guez}, E. and {Arellano Ferro}, A. and {Alvarez}, M. and {Parrao}, L. and {Pe{\~n}a}, J. and {Eyer}, L. and {Aerts}, C. and {De Cat}, P. and {Weiss}, W.~W. and {Zhou}, A.},
        title = "{Multi-site, multi-technique survey of {\ensuremath{\gamma}} Doradus candidates. I. Spectroscopic results for 59 stars}",
      journal = {\aap},
         year = 2004,
        month = apr,
       volume = {417},
        pages = {189-199},
          doi = {10.1051/0004-6361:20034503},
archivePrefix = {arXiv},
       eprint = {astro-ph/0402397},
 primaryClass = {astro-ph},
       adsurl = {https://ui.adsabs.harvard.edu/abs/2004A&A...417..189M}
}

@ARTICLE{Schrijvers1999,
       author = {{Schrijvers}, C. and {Telting}, J.~H.},
        title = "{Line-profile variations due to adiabatic non-radial pulsations in rotating stars. IV. The effects of intrinsic profile variations on the IPS diagnostics}",
      journal = {\aap},
         year = 1999,
        month = feb,
       volume = {342},
        pages = {453-463},
       adsurl = {https://ui.adsabs.harvard.edu/abs/1999A&A...342..453S}
}

@ARTICLE{Murphy2019,
       author = {{Murphy}, Simon J. and {Hey}, Daniel and {Van Reeth}, Timothy and {Bedding}, Timothy R.},
        title = "{Gaia-derived luminosities of Kepler A/F stars and the pulsator fraction across the {\ensuremath{\delta}} Scuti instability strip}",
      journal = {\mnras},
         year = 2019,
        month = may,
       volume = {485},
       number = {2},
        pages = {2380-2400},
          doi = {10.1093/mnras/stz590},
archivePrefix = {arXiv},
       eprint = {1903.00015},
 primaryClass = {astro-ph.SR},
       adsurl = {https://ui.adsabs.harvard.edu/abs/2019MNRAS.485.2380M}
}

@ARTICLE{Skarka2022,
       author = {{Skarka}, M. and {{\v{Z}}{\'a}k}, J. and {Fedurco}, M. and {Paunzen}, E. and {Henzl}, Z. and {Ma{\v{s}}ek}, M. and {Karjalainen}, R. and {Sanchez Arias}, J.~P. and {S{\'o}dor}, {\'A}. and {Auer}, R.~F. and {Kab{\'a}th}, P. and {Karjalainen}, M. and {Li{\v{s}}ka}, J. and {{\v{S}}tegner}, D.},
        title = "{Periodic variable A-F spectral type stars in the northern TESS continuous viewing zone. I. Identification and classification}",
      journal = {\aap},
         year = 2022,
        month = oct,
       volume = {666},
          eid = {A142},
        pages = {A142},
          doi = {10.1051/0004-6361/202244037},
archivePrefix = {arXiv},
       eprint = {2207.12922},
 primaryClass = {astro-ph.SR},
       adsurl = {https://ui.adsabs.harvard.edu/abs/2022A&A...666A.142S}
}

@ARTICLE{Rice2023,
       author = {{Rice}, Malena and {Wang}, Songhu and {Gerbig}, Konstantin and {Wang}, Xian-Yu and {Dai}, Fei and {Tyler}, Dakotah and {Isaacson}, Howard and {Howard}, Andrew W.},
        title = "{The Orbital Architecture of Qatar-6: A Fully Aligned Three-body System?}",
      journal = {\aj},
         year = 2023,
        month = feb,
       volume = {165},
       number = {2},
          eid = {65},
        pages = {65},
          doi = {10.3847/1538-3881/aca88e},
archivePrefix = {arXiv},
       eprint = {2212.02542},
 primaryClass = {astro-ph.EP},
       adsurl = {https://ui.adsabs.harvard.edu/abs/2023AJ....165...65R}
}

@ARTICLE{Penev2018,
       author = {{Penev}, Kaloyan and {Bouma}, L.~G. and {Winn}, Joshua N. and {Hartman}, Joel D.},
        title = "{Empirical Tidal Dissipation in Exoplanet Hosts From Tidal Spin-up}",
      journal = {\aj},
         year = 2018,
        month = apr,
       volume = {155},
       number = {4},
          eid = {165},
        pages = {165},
          doi = {10.3847/1538-3881/aaaf71},
archivePrefix = {arXiv},
       eprint = {1802.05269},
 primaryClass = {astro-ph.SR},
       adsurl = {https://ui.adsabs.harvard.edu/abs/2018AJ....155..165P}
}

@ARTICLE{Naoz2016,
       author = {{Naoz}, Smadar},
        title = "{The Eccentric Kozai-Lidov Effect and Its Applications}",
      journal = {\araa},
         year = 2016,
        month = sep,
       volume = {54},
        pages = {441-489},
          doi = {10.1146/annurev-astro-081915-023315},
archivePrefix = {arXiv},
       eprint = {1601.07175},
 primaryClass = {astro-ph.EP},
       adsurl = {https://ui.adsabs.harvard.edu/abs/2016ARA&A..54..441N}
}

@ARTICLE{Kuffmeier2024,
       author = {{Kuffmeier}, M. and {Pineda}, J.~E. and {Segura-Cox}, D. and {Haugb{\o}lle}, T.},
        title = "{Constraints on the primordial misalignment of star-disk systems}",
      journal = {\aap},
         year = 2024,
        month = oct,
       volume = {690},
          eid = {A297},
        pages = {A297},
          doi = {10.1051/0004-6361/202450410},
archivePrefix = {arXiv},
       eprint = {2405.12670},
 primaryClass = {astro-ph.SR},
       adsurl = {https://ui.adsabs.harvard.edu/abs/2024A&A...690A.297K}
}

@article{Hwang2020,
    author = {Hwang, Hsiang-Chih and Hamer, Jacob H and Zakamska, Nadia L and Schlaufman, Kevin C},
    title = {Very wide companion fraction from Gaia DR2: A weak or no enhancement for hot Jupiter hosts, and a strong enhancement for contact binaries},
    journal = {Monthly Notices of the Royal Astronomical Society},
    volume = {497},
    number = {2},
    pages = {2250-2259},
    year = {2020},
    month = {07},
    issn = {0035-8711},
    doi = {10.1093/mnras/staa2124},
    url = {https://doi.org/10.1093/mnras/staa2124},
    eprint = {https://academic.oup.com/mnras/article-pdf/497/2/2250/33569358/staa2124.pdf},
}

@ARTICLE{Yang2025,
       author = {{Yang}, Eritas and {Su}, Yubo and {Winn}, Joshua N.},
        title = "{A Third Star in the HAT-P-7 System and a New Dynamical Pathway to Misaligned Hot Jupiters}",
      journal = {\apj},
         year = 2025,
        month = jun,
       volume = {986},
       number = {2},
          eid = {117},
        pages = {117},
          doi = {10.3847/1538-4357/add5f7},
archivePrefix = {arXiv},
       eprint = {2505.07927},
 primaryClass = {astro-ph.EP},
       adsurl = {https://ui.adsabs.harvard.edu/abs/2025ApJ...986..117Y}
}
\bibliographystyle{aasjournalv7}

\appendix
\renewcommand{\thefigure}{A\arabic{figure}} 
\setcounter{figure}{0}

\begin{figure}[h]
    \centering
    \includegraphics[width=\linewidth]{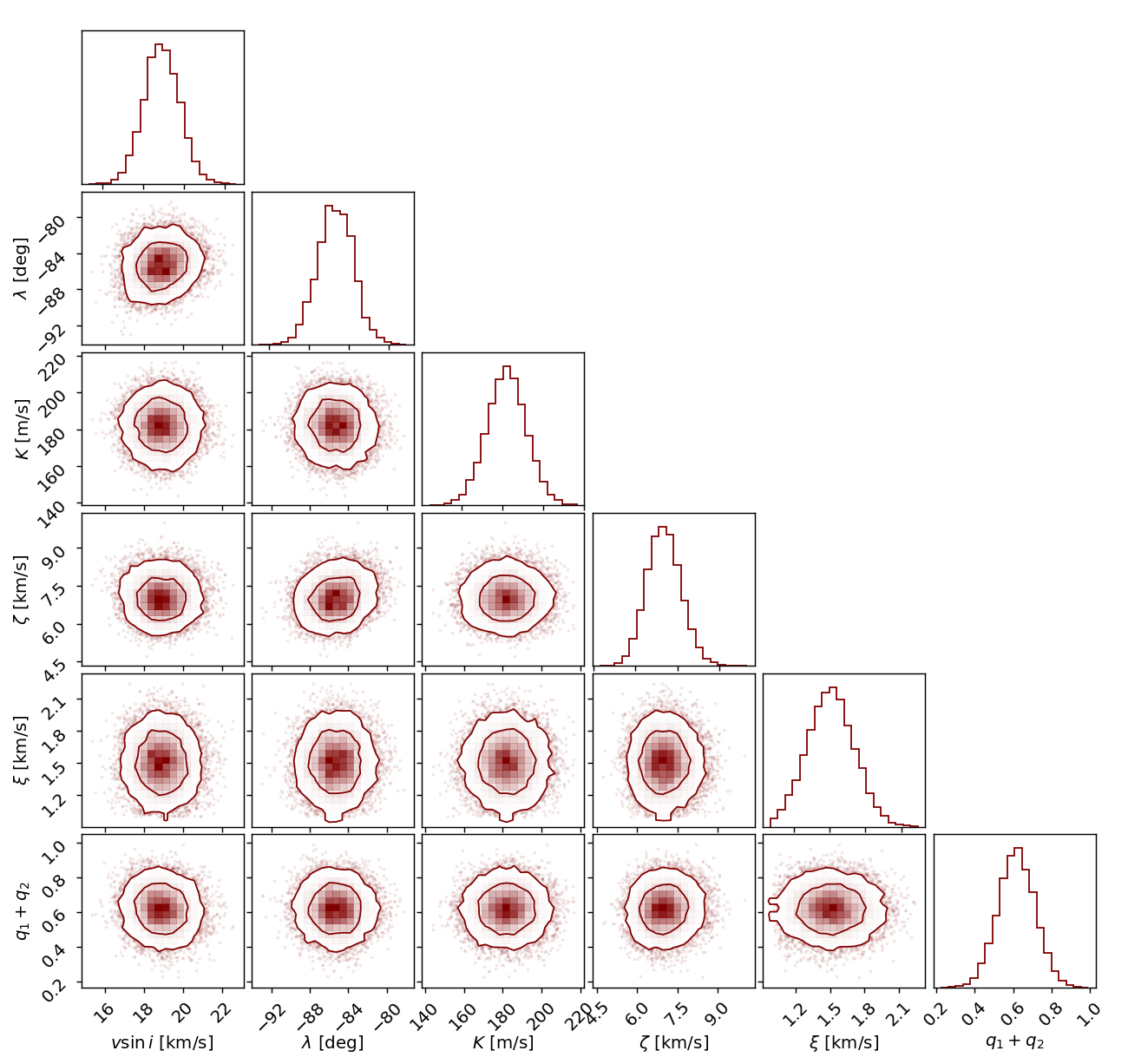}
    \caption{Best-fit corner plot from fitting HAT-P-49 displays the mutual correlations between all fitted parameters, and their posterior distributions. The contours represent the 1 and 2\,$\sigma$ confidence regions.}
    \label{fig:best_corner_h49}
\end{figure}

\begin{figure}[b]
    \centering
    \includegraphics[width=\linewidth]{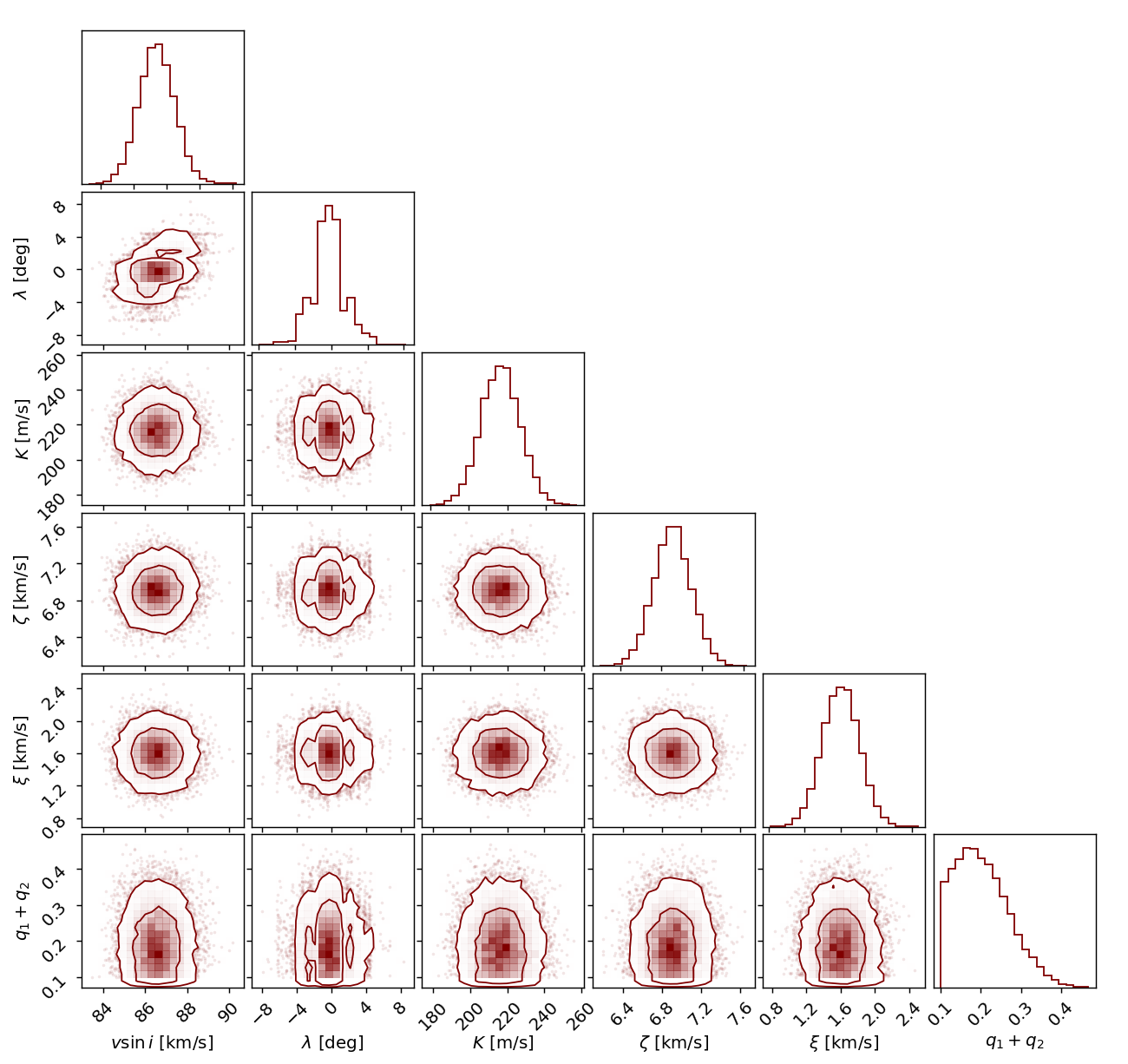}
    \caption{Best-fit corner plot from fitting HAT-P-57\,A.}
    \label{fig:best_corner_h57}
\end{figure}

\begin{figure}[b]
    \centering
    \includegraphics[width=\linewidth]{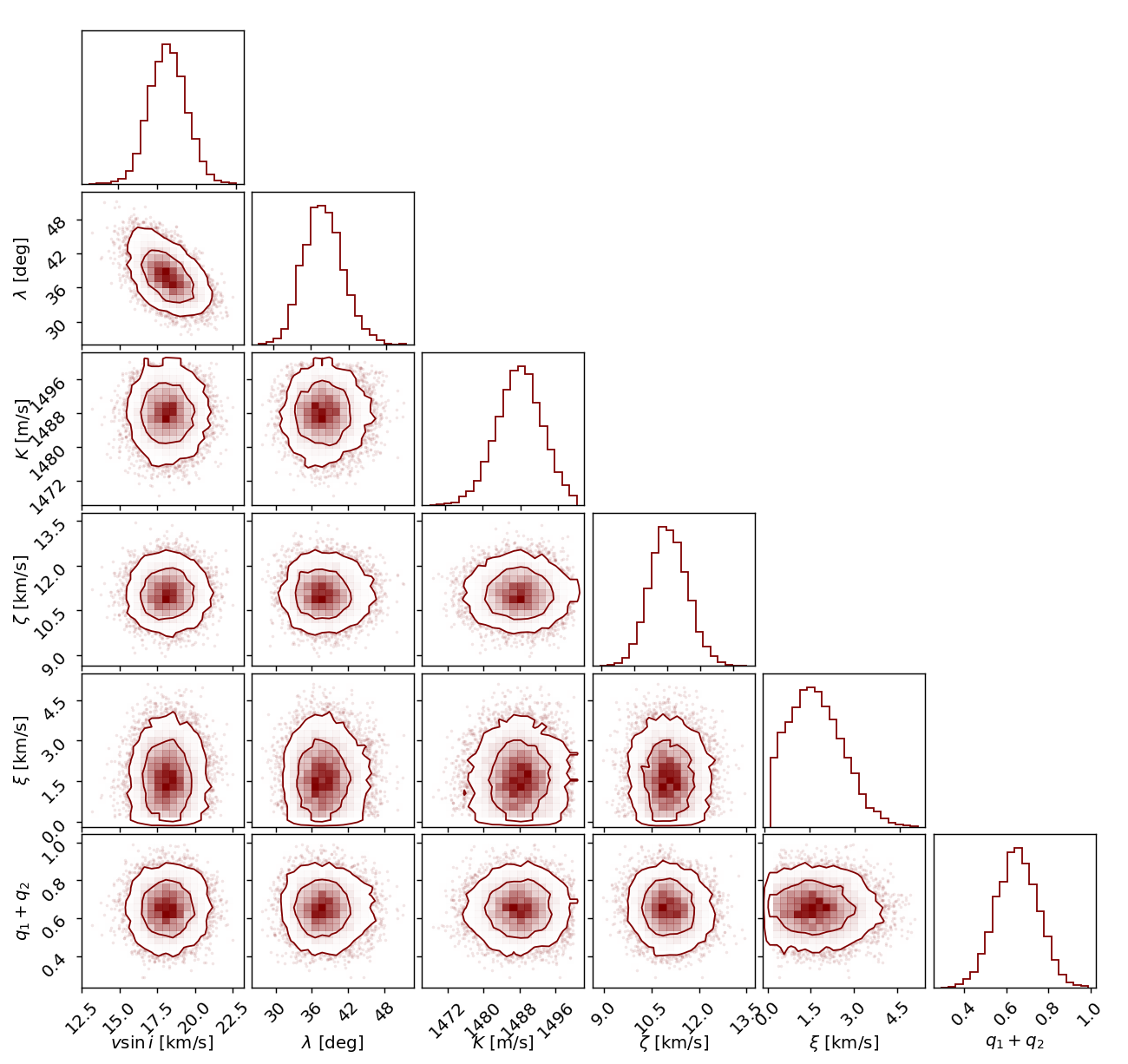}
    \caption{Best-fit corner plot from fitting XO-3\,A.}
    \label{fig:best_corner_xo3}
\end{figure}




\end{document}